\documentclass[reprint,floatfix,superscriptaddress]{revtex4-2}

\usepackage{amsmath}
\usepackage{amssymb}
\usepackage{graphicx}
\usepackage{dcolumn}
\usepackage{bm}
\usepackage{xcolor}
\usepackage{multirow}
\usepackage{booktabs}
\usepackage{float}
\usepackage{tabularx}
\begin{document}


\title{Direct laser writing for cardiac tissue engineering:  a microfluidic heart on a chip with integrated transducers}

\author{Rachael K. Jayne}
\altaffiliation{These authors contributed equally to this work.}
\affiliation{Department of Mechanical Engineering, Boston University, Boston, MA 02215, USA}
\affiliation{Photonics Center, Boston University, Boston, MA 02215, USA}
\author{M. Çağatay Karakan}
\altaffiliation{These authors contributed equally to this work.}
\affiliation{Department of Mechanical Engineering, Boston University, Boston, MA 02215, USA}
\affiliation{Photonics Center, Boston University, Boston, MA 02215, USA}
\author{Kehan Zhang}
\altaffiliation{These authors contributed equally to this work.}
\affiliation{Department of Biomedical Engineering, Boston University, Boston, MA 02215, USA}
\affiliation{Wyss Institute for Biologically Inspired Engineering, Harvard University, Boston, MA 02115, USA}
\author{Noelle Pierce}
\affiliation{Photonics Center, Boston University, Boston, MA 02215, USA}
\author{Christos Michas}
\affiliation{Department of Biomedical Engineering, Boston University, Boston, MA 02215, USA}
\author{David J. Bishop}
\affiliation{Department of Mechanical Engineering, Boston University, Boston, MA 02215, USA}
\affiliation{Department of Biomedical Engineering, Boston University, Boston, MA 02215, USA}
\affiliation{Department of Electrical and Computer Engineering, Boston University, Boston, MA 02215, USA}
\affiliation{Division of Materials Science and Engineering, Boston University, Boston, Massachusetts 02215, USA}
\affiliation{Department of Physics, Boston University, Boston, MA 02215, USA}
\author{Christopher S. Chen}
\affiliation{Department of Biomedical Engineering, Boston University, Boston, MA 02215, USA}
\affiliation{Wyss Institute for Biologically Inspired Engineering, Harvard University, Boston, MA 02115, USA}
\author{Kamil L. Ekinci}
\email[Electronic mail:]{ekinci@bu.edu}
\affiliation{Department of Mechanical Engineering, Boston University, Boston, MA 02215, USA}
\affiliation{Photonics Center, Boston University, Boston, MA 02215, USA}
\affiliation{Division of Materials Science and Engineering, Boston University, Boston, Massachusetts 02215, USA}
\author{Alice E. White}
\email[Electronic mail:]{aew1@bu.edu}
\affiliation{Department of Mechanical Engineering, Boston University, Boston, MA 02215, USA}
\affiliation{Photonics Center, Boston University, Boston, MA 02215, USA}
\affiliation{Division of Materials Science and Engineering, Boston University, Boston, Massachusetts 02215, USA}
\affiliation{Department of Physics, Boston University, Boston, MA 02215, USA}
\affiliation{Department of Electrical and Computer Engineering, Boston University, Boston, MA 02215, USA}

\date{\today}

\begin{abstract}
We have designed and fabricated a microfluidic-based platform for sensing mechanical forces generated by cardiac microtissues in a highly-controlled microenvironment. Our fabrication approach combines Direct Laser Writing (DLW) lithography with soft lithography. At the center of our platform is a cylindrical volume, divided into two chambers by a cylindrical polydimethylsiloxane (PDMS) shell. Cells are seeded into the inner chamber from a top opening, and the microtissue assembles onto tailor-made attachment sites on the inner walls of the cylindrical shell. The outer chamber is electrically and fluidically isolated from the inner one by the cylindrical shell and is designed for actuation and sensing purposes.  Externally applied pressure waves to the outer chamber deform parts of the cylindrical shell and thus allow us to exert time-dependent forces on the microtissue. Oscillatory forces generated by the microtissue similarly deform the cylindrical shell and change the volume of the outer chamber, resulting in measurable electrical conductance changes. We have used this platform to study the response of cardiac microtissues derived from human induced pluripotent stem cells (hiPSC) under prescribed mechanical loading and pacing.
\end{abstract}

\maketitle

\section{Introduction}

The generation of microscale engineered cardiac tissues, also known as heart-on-a-chip systems, for studying heart physiology and disease has advanced substantially in recent years \cite{weinberger2017engineering,zhang2018can,kitsara2019heart}. Heart-on-a-chip systems provide environments that mimic native tissue and typically allow for some control over  the relevant parameter space. Recent studies employing these platforms  have given researchers invaluable  insight into the biology of human cardiac tissue  and  allowed for some  high-throughput testing  \cite{Boudou2012,Hinson2015,ronaldson2018advanced}. A number of these platforms, such as muscular thin films,  have predominantly focused on 2D microtissues. Laminar cardiac microtissues with embedded strain gauges have allowed for  studies  of contractile stresses inside the microtissues  \cite{lind2017cardiac,lind2017instrumented} and are well suited for high-throughput drug screening but do not allow for application of mechanical forces  or strains to the microtissue. This shortcoming has been addressed in both commercial  and custom-made platforms fitted with a variety of actuators, in which static or dynamic strains can readily be applied to 2D monolayers of cells\cite{flexcell,strexcell,Kreutzer2014}.  Some studies in  these platforms have hinted that application of strains  enhance levels of functionality and maturation in cardiac tissue monolayers; however, cell-cell or cell-extracellular matrix interactions present in 3D have naturally been left out of the picture in 2D microtissues. The  3D morphology of human cardiac muscle tissue has been more closely mimicked in a 3D cardiac tissue platform featuring deformable polydimethylsiloxane (PDMS) pillars \cite{abilez2018passive,Legant2009,Boudou2012,Galie2015}.  These pillar structures constrain and guide the cardiac cells and collagen gel into freestanding tissue constructs. By tuning the mechanical properties of  PDMS micropillars and the gel encapsulating the entire structure, significant variations in tissue morphology have been achieved. Several studies have also shown that tuning the stiffness of the substrates significantly alters the static and dynamic tension generated by microtissues \cite{Boudou2012,ma2018contractile,guo2020elastomer}. There have also been some efforts directed to integrating mechanical actuators \cite{javor2020microtissue}, such as pneumatic actuators \cite{Parsa2017,furuike2018pneumatically,marsano2016beating}, into these micropillar platforms to study the effects of mechanical stimuli. These studies  have  shown that mechanical actuation can facilitate tissue functional maturation or induce diseases, such as cardiac hypertrophy\cite{fink2000chronic}.

\begin{figure*}[t]
	\includegraphics[width=\linewidth]{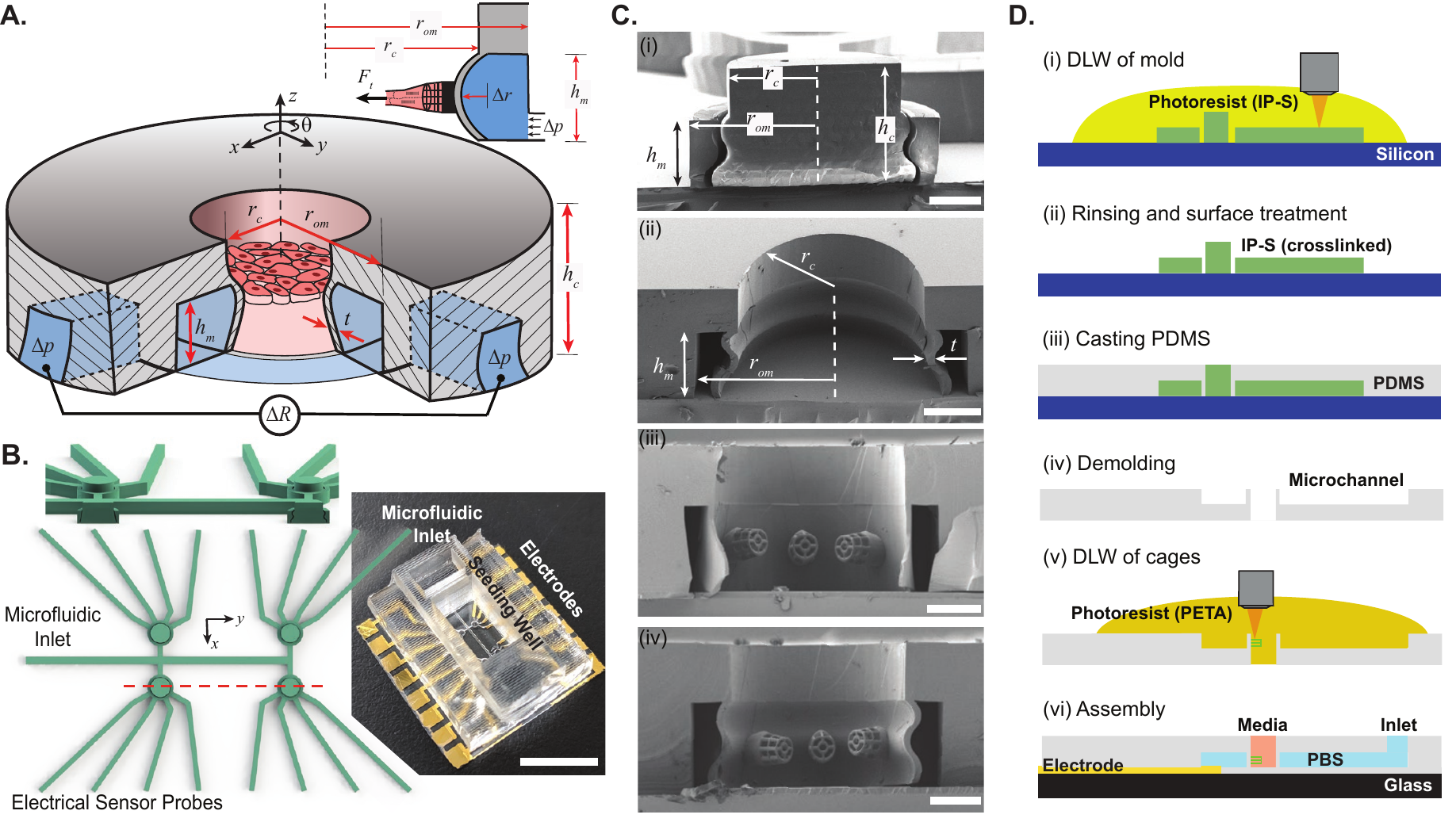}
	\caption[Test system fabrication process]{Overview and fabrication of the platform. (A) Illustration of the platform. The cut shows the inner cylindrical well  (i.e., the seeding well) for growing suspended cardiac microtissues and the annular microchannel for actuation and sensing. The externally applied pressure $\Delta p$ is transduced into a  strain by the bending of the thin microchannel wall. An electrical resistance measurement through the microchannel ($\Delta R$) allows for force sensing. The inset shows  a cross-section view of the bulging wall of the seeding well, highlighting the tailor-made cell attachment microstructures.   (B)  The 3D mold design. The top inset shows a cross-sectional view through the seeding wells (dotted line). The photograph on the right shows a completed platform. Individual device cavities and microfluidic channels are fluidically isolated. The larger top opening (seeding well) provides access to the individual seeding wells of the devices  for pipetting the cells whereas the microfluidic inlet is connected to the annular microchannels. The scale bar is 10 mm. (C) SEM images of devices at various points during the fabrication. The  structures were cut in the middle in order to show their important features. (i) Mold of a device with curved walls; (ii) the PDMS structure made from this mold; (iii) cell attachment structures on planar  and (iv) curved walls. The scale bar is 200 $\mu$m. (D) Fabrication steps. (i-ii) Negative master molds are fabricated via DLW lithography on silicon substrates. (iii-iv) PDMS is cast into these molds and demolded. (v) A second DLW step results in the microstructures for cell attachment on the inner walls of the cavities. (vi) Devices are then bonded to an electrode-patterned substrate  and seeded with cells.} 
	\label{fig:fig1}
\end{figure*}

While different actuation techniques have been explored in existing devices, optical microscopy has  been the favored technique  for studying the mechanical responses of the  cardiac microtissues. Optical techniques have certain advantages: for example, they allow for the measurement of the electrophysiological characteristics of the microtissues in conjunction with their contractile displacements \cite{mathur2015human,wang2019biowire}. However, they also come with some shortcomings. First, there are  challenges associated with imaging multiple  devices (or microtissues) in parallel for high-throughput studies. Second, post processing (i.e.,  image analysis) is typically required, making real time measurements difficult. Finally, the need for a microscope equipped with an incubator complicates experimental set ups. 

Here, we present a platform that builds on the  desirable aspects of earlier devices and addresses some of their shortcomings. The platform may serve as a multi-functional and scalable toolbox for cardiac tissue engineering and enable: (i) 3D self-assembly and growth of cardiac tissue in customizable geometries and orientations \cite{naito2006optimizing,bose2018effects}, (ii) real-time and parallel detection of contractile stresses exerted by multiple microtissues, (iii) precise and dynamic control of  external mechanical cues. Our platform shown in Fig. \ref{fig:fig1}A and B  consists of an array of four devices with microfluidic actuators  and  integrated electrical sensors. Each individual device  has a microtissue seeding well at its center with an embedding actuator  \cite{Kreutzer2014}. The integrated sensors provide  electronic readout of tissue contractile stresses under prescribed forces (strains) in real time. Each device allows for culturing, and subsequently experimenting on, a 3D  freestanding  cardiac microtissue with controlled  alignment and geometry. The tissue alignment is  enabled by tailor-made adhesion sites on the walls of the seeding well. In order to show the unique capabilities of this platform, we have studied  the mechanical properties of cardiac microtissues derived from human induced pluriopotent stem cells (hiPSC). In particular, we have observed an increase in active contractile forces from the microtissues with increasing tissue length, consistent with the fundamental force–length (Frank-Starling) relationship of cardiac muscle. Further, we have attempted to mechanically entrain a cardiac microtissue by periodically modulating the applied strain \cite{Nitsan2016}. In this experiment, we have observed a brief synchronization between the spontaneous beating  and the externally applied mechanical perturbation. In the near future, this platform may allow for high-throughput and real-time investigations of 3D cardiac microtissue maturation under complex mechanical, electrical, and chemical conditioning. 

\section{Results}

\subsection{Overall Device Design and Operation}

Each PDMS platform has a  $2\times 2$ array of  microfluidic heart-on-a-chip devices. Fig. \ref{fig:fig1}A is an illustration of  a single device,  with the cut showing the  seeding well, the microtissue, and the annular microfluidic channel.  Fig. 1B shows computer-aided drawings of the platform mold  and a photograph of a completed platform. Each device  (Fig. \ref{fig:fig1}A) is based on a  cylindrical cavity at its center. These cylindrical cavities of radii $r_{c} \approx \rm 400~ \rm \mu m$ and height $h_c\approx 500 ~\rm \mu m$ act as seeding wells and are open on the top side. Each seeding well is surrounded by an annular  microfluidic channel. The height  of this microfludic channel is $h_m \approx 300~ \rm \mu m$; the seeding wells and the surrounding microfluidic channels   are  separated by a thin  compliant cylindrical shell of thickness $t\approx \rm 20-30~ \mu m$; the inner  radius $r_{im}$ of the microchannel depends on the shell deflection (Fig. 1A inset) but, initially, is $r_{im}\approx r_{c}+t \approx 420-430~\rm \mu m$; the outer radius $r_{om} \approx 480-540~\rm \mu m$. The top of this shell is anchored to a  bulk PDMS piece of thickness $\approx \rm 200~ \mu m$, and the bottom is bonded to a glass substrate. All the linear dimensions can  be seen in the illustration in Fig. 1A and the SEM images in Fig. \ref{fig:fig1}C taken at different points during the fabrication process.

The  annular microchannel surrounding the central seeding well enables actuation and sensing  (Fig. \ref{fig:fig1}A and C). The fluid inlets of all four devices on the platform are routed to a common inlet that is connected to a microfluidic pump (Fig. \ref{fig:fig1}B). The pump fills the devices with a solution of $1\times$ phosphate buffered saline (PBS) and controls the pressure in the annular microchannels. A positive or negative pressure differential $\Delta p$ applied between the  seeding well and the annular microchannel can control both the direction and the amplitude of the bending of the  PDMS shell, which results in prescribed strains  on the microtissues.  Change in the volume, and hence the electrical resistance $\Delta R$, of the  annular channel is proportional to the shell displacement, providing a sensitive sensing mechanism.

Two of the four seeding wells on a single platform are surrounded by 20-$\mu$m-thick planar shells (Fig. \ref{fig:fig1}C(iii)) and the other two are surrounded by 30-$\mu$m-thick ``curved'' shells (Fig. \ref{fig:fig1}C(iv)). We found that the 20-$\mu$m-thick shells were robust enough to withstand demolding and stiff enough to maintain their shape afterward. The  curved shell structure  exploited the unique advantage of the direct laser writing (DLW) in the fabrication process, providing more mechanical robustness compared to the planar shell structure. The 30-$\mu$m-thick curved shells are mechanically less compliant than the 20-$\mu$m-thick planar shells. The curved geometry also tends to distribute the deformation more uniformly across the cylindrical surface based on finite element simulations, see Electronic Supplementary Information (ESI) for details. A slight 10-degree inward taper was added to the shells to prevent adhesion problems during the second DLW step in which polymer  microstructures in the form of cages were printed on the sides of the inner walls of the shell (Fig. \ref{fig:fig1}C(iii)-(iv)). These cages are an important and unique feature of our devices: they provide tailor-made  locations for cardiac tissue attachment and allow for control of the tissue geometry.

After cells are seeded into the devices and centrifuged, they self-assemble onto the attachment sites, compact over several days, and form suspended 3D tissue constructs between attachment sites. Walls of the seeding wells deflect as a result of either externally applied pressure differentials or tissue-generated forces. As discussed below, the measured change in electrical resistance of the microchannels can be calibrated, making it possible to monitor the forces applied on the PDMS shell in real time without relying on optical imaging.

\begin{figure*}[t]
	\centering
	\includegraphics[width=\linewidth]{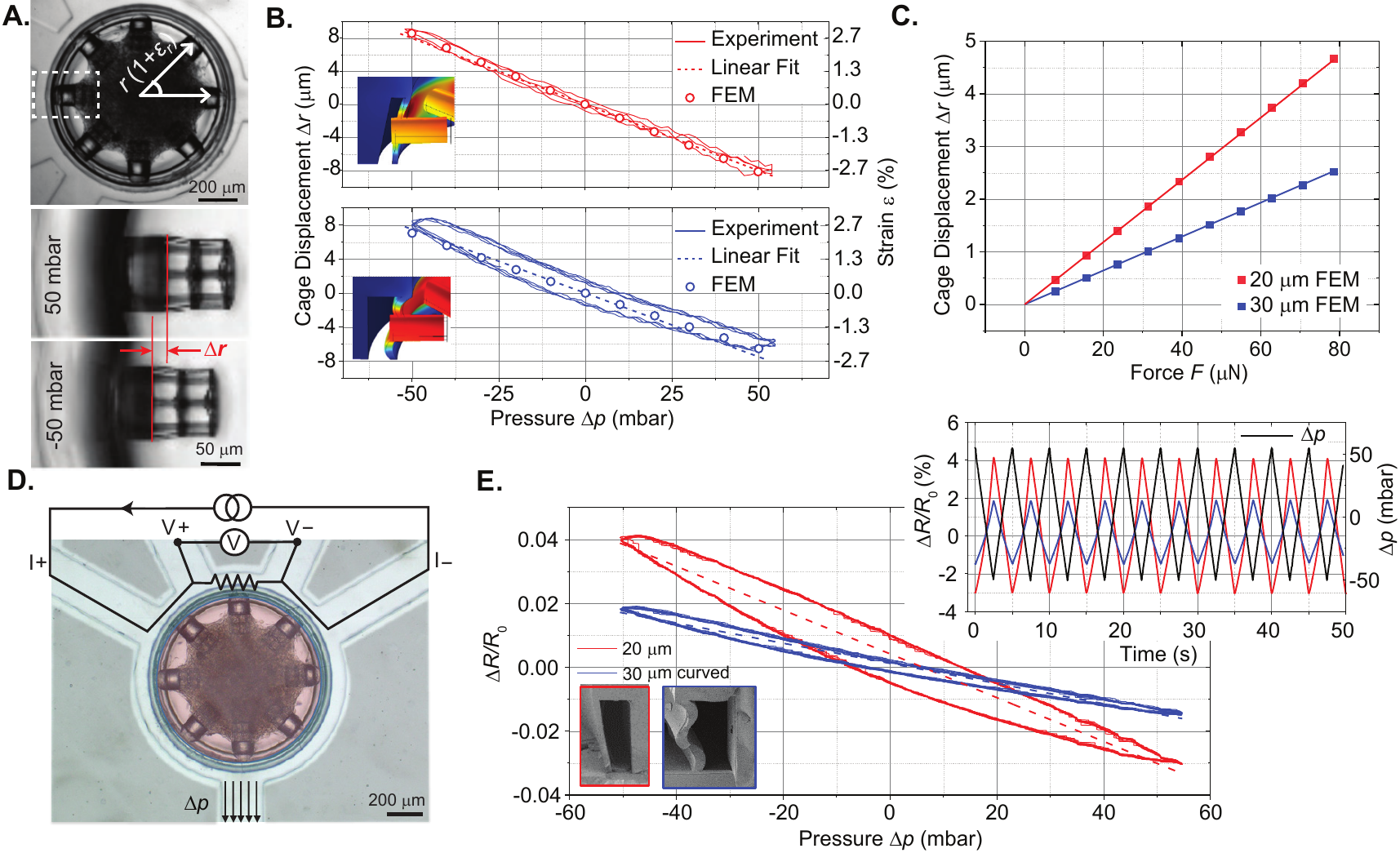}
	\caption{Calibration of the platform. (A) Optical image (top) of the 30-$\mu$m device (with a cardiac microtissue inside) showing the  radial strain on the cylindrical shell. On the bottom, there are close up  images of a cell-attachment site ``cage", showing how the optical calibration was performed. The cage stubs were tracked as different pressures $\Delta p$ were applied. The two images are at $\Delta p= 50$ and $-50$ mbar, and an increase in  radius of $\Delta r \approx 16 ~ \rm \mu m$ is measured at $z=h_{m}/2$ (B) $\Delta r$  as a function of $\Delta p$ for the 20-$\mu$m-thick planar (red, top) and 30-$\mu$m-thick curved (blue, bottom) devices on the same platform. Solid lines show the optical data collected with an inverted microscope after image processing. Dashed lines represent simple linear fits: $\Delta r=0.16 \Delta p$ (top), $\Delta r=0.15 \Delta p$ (bottom). Open symbols are from finite element simulations. (C) FEM results (symbols) showing  cage displacement as a function of  the force applied on the cages. Solid lines are fits to obtain the  spring constants, $k_{eff}=16.9$ N/m (20-$\mu$m planar) and $k_{eff}=31.1$ N/m (30-$\mu$m curved) (D) False-coloured image showing the top view of a device with connecting microchannels:  seeding well  with cardiac microtissue and cell media (red), and annular microchannel and connecting microchannels (light blue). The top region of the annular microchannel acts as the electrical sensor, with the sensing performed by the four-wire resistance measurement circuit.   A pressure $\Delta p$  (bottom) results in a  strain. Scale bar is 200 $\mu$m. (E) Relative resistance changes of the sensor measured as a function of the applied pressure (10 cycles each). The dashed lines are linear fits providing the electrical responsivity ${\cal R}_e$. The top inset shows how the measurement was taken by applying a periodic triangular $\Delta p$ at 0.2 Hz as a function of time  and measuring $\Delta R/R_0$. The left inset shows the SEM image of PDMS shell and microchannel cross sections.}
	\label{fig:fig2}
\end{figure*}

\subsection{Fabrication Approach}
We used a combination of soft lithography and DLW for the fabrication process. Advantages offered by DLW allowed us to fabricate the key features of the platform. DLW  generates features in a liquid photoresist by tightly focusing a femtosecond laser in the photoresist to induce two-photon polymerization (2PP). By scanning the laser spot and using a 3-axis piezo stage, one can define and print 3D structures in the photoresist coated on various substrates with sub-micron resolution \cite{song2020simple}. In this study, a commercial DLW system (Nanoscribe Photonic Professional GT) with a $25 \times$ (NA=0.8) immersion objective is used, which provides a minimum resolution of diameter $0.6~\rm \mu m$ in the horizontal plane and height $1.5~ \rm \mu m$ along the vertical axis.

Fabrication steps of the platform are shown in  Fig. \ref{fig:fig1}D. Briefly, negative master molds are printed using DLW (Fig. D(i)) on a negative tone proprietary photoresist (IP-S, $E_{\rm IP-S}= 4.6 ~ \rm GPa$), which is drop-cast on a silicon substrate. The mold design is shown in Fig. \ref{fig:fig1}B, and  the seeding well region of a completed mold is shown in Fig. \ref{fig:fig1}C(i).  The negative master molds fabricated by DLW have two important features that cannot be obtained in traditional lithography: structures with curves along the vertical axis, and truly 3D structures with different heights on a plane. Once negative master molds are printed and excess photoresist is rinsed away, molds are fluorinated to prevent stiction (Fig 1D(ii). PDMS is cast onto the molds (Fig. 1D (iii)), sandwiched, cured, and demolded  (Fig. 1D(iv)), resulting in a $0.5$-mm-thick PDMS device layer with embedded microfluidic channels and open seeding wells. The SEM image in Fig. 1C(ii) shows a device after the demolding step. The demolded PDMS device undergoes another step of DLW in which micron-scale structures  are printed on the sides of the wells (Fig. 1D(v)). The photoresist used in this step is pentaerythritol triacrylate (PETA) with 3 wt$\%$ Irgacure 819 (BASF) photoinitiator, which is $\sim 100\times$ harder than PDMS ($E_{\rm PETA}\approx 260 ~ \rm MPa$\cite{jayne2018}) and electrically insulating. These microstructures; shown in the SEM images in Fig. 1C(iii)-(iv), look like cages and facilitate cell attachment.  Finally, the devices are sealed by bonding the PDMS layer to a glass substrate with metal electrodes (Fig. 1D (vi)); subsequently, a thicker  PDMS piece that contains a media reservoir and a PBS well is bonded on top of the entire platform. Figure \ref{fig:fig1}B shows the photograph of a completed platform ready for cell seeding and testing. 

\subsection{Device Calibration}

In cardiac microtissue testing, we typically apply constant or oscillatory strains to the microtissue and monitor its time-dependent contractions. As mentioned above, the strain is imposed by  applying a pressure $\Delta p$ to the fluid in the annular chamber, which is transduced into a displacement  by the bending of the enclosure shell  (Fig. \ref{fig:fig1}A inset). We assume that  any disturbance on the wall is propagated to the microtissue without attenuation. This assumption is based on the observation that the cage microstructures are mechanically much stiffer compared to both the walls and the microtissue (ESI). The first leg of our calibration  involves determining the mechanical responsivity of this displacement transducer; the force constant that the tissue experiences is also related to this calibration. The second leg of our calibration is for determining the electrical responsivity of our sensor, which converts  the wall displacement into electrical resistance changes. 

Figure \ref{fig:fig2}A-B shows how  the applied $\Delta p$ is converted  into a strain $\varepsilon_r$ using optical images. The radial displacement $\Delta r$ of the cage stub at its center point,  $z= h_m/2$, is measured in an inverted microscope as a function of the applied $\Delta p$ (see Movie S1). (Since the cage is much stiffer compared to the shell, we neglect its deformation in the analysis). The data traces in Fig. 2B (continuous lines) show the experimentally measured $\Delta r$ as a function of $\Delta p$ for the 20-$\mu$m planar (top) and 30-$\mu$m curved (bottom) shells. The corresponding strain values (right $y$ axes) are calculated from $\varepsilon_r \approx \Delta r/r_{c}$, assuming cylindrical symmetry. The data follow a mostly linear trend  (small deviations from linearity are discussed in ESI), allowing us to determine  a linear mechanical responsivity for the transducer as, ${\cal R}_m = \frac{\partial \varepsilon_r}{\partial \Delta p}$, from the fit. The dotted lines in Fig. 2B are linear fits.  We  also compare these measurements with results from Finite Element Models (FEM) shown by data points (open circles) in Fig. 2B. Here, we have modeled both types of devices with the 20-$\mu$m-thick planar and 30-$\mu$m-thick curved shells. With the modulus of the PDMS shell taken as   $E_{PDMS} \approx2.2~ \rm MPa$   \cite{Johnston2014} and the modulus of pentaerythritol triacrylate (PETA) cages taken    as $E_{PETA} \approx 260~\rm MPa$ \cite{jayne2018}, the FEM simulations  are in excellent agreement with the experimental measurements. The  agreement suggests that our FEM simulations provide an accurate description of the system and encourages us to rely on FEM for further characterization of the system.  More details of the FEM simulations are available in the ESI.

Conversely, we can determine the effective spring constant, $k_{eff} \approx \frac{\partial F}{\partial r_c}$, that the walls present to forces applied on the cage microstructures.   This $k_{eff}$ informs us of the time-dependent forces exerted by the microtissue. Encouraged by the above-mentioned FEM models, we have applied a uniform normal stress on the stubs of the cages in the $- {r}$ direction  in $10$ mbar increments (see ESI) to model tissue forces.  Next, we have extracted  the resulting radial displacement of the cage at $z= h_m/2$  as a function of the applied stress. We have assumed that the cages are rigid, as above. Figure \ref{fig:fig2}C shows the results of this calibration. Here, the $x$-axis is in units of the force obtained  by multiplying the stress by the appropriate cage area.  The linear relation between force and the  radial displacement provides the spring constant ${k_{eff}} \approx {\frac{\partial F}{\partial \Delta r}}$.  The experimentally determined values  are ${k_{eff}} \approx 16.9 ~\rm N/m$ and ${k_{eff}} \approx 31.1 ~\rm N/m$ for the 20-$\mu$m and 30-$\mu$m-thick transducers, respectively. At a first glance, one may think that  ${k_{eff}}$ and  ${\cal R}_m $ can be related as ${k_{eff}} \sim {\frac{S}{{r_{c}{\cal R}_m}}}$, where $S \approx 2\pi {r_c} h_m$ represents the  area over which the externally applied  $\Delta p$ acts. However, the complex geometry of the structure results in different bending patterns (and different $\Delta r$) for pressures applied uniformly from the outside as compared to forces applied on the cage microstructures from the inside of the seeding well. Given the complex bending patterns (ESI), using $\Delta r$ as the relevant experimental parameter may be an oversimplification, although providing a practical and linear description.    

Finally, a unique feature of the device is that it allows all-electrical  sensing of the contractile forces exerted  by the microtissue. In fact, any deformation of the seeding well walls, such as those generated by  an externally  applied  pressure, are detectable electrically. The electrical sensing principle and circuit are shown in Fig. 2D. The part of the microchannel wrapped around the seeding well (shaded in the Fig. \ref{fig:fig2}D) forms the sensing resistor.  Assuming a rectangular cross-section and a small curvature for the moment to illustrate the sensing principle, the initial resistance is $R_0 \propto \frac{1}{h(r_{om}-r_{im})}$, where $r_{im}$  and $r_{om}$ are respectively the inner and outer radii of the microchnanel (Fig. \ref{fig:fig1}A).  As $r_{im}$ changes due to forces applied to the shell, the electrical resistance also changes. The principle is similar to that of a microfluidic strain gauge, but with the width of the microchannel changing rather than the length.  We perform the electrical calibration by measuring the electrical resistance change as a function of the applied pressure as shown in Fig. \ref{fig:fig2}E. The inset of Fig. 2E shows the triangular pressure waveform applied to the device during the resistance measurement. The planar device (red curve) responds with a larger resistance change and appears more hysteretic compared to the curved device (blue curve). Regardless, a linear fit (dotted lines in Fig. 2E) to both provides a sufficiently high-fidelity description of the electrical measurement.  The  linear fits yield  ${\cal R}_e = \frac {\partial \Delta R}{R_0 \partial \Delta p}$. We can further show that \cite{ozsun2013}, to first order, $\frac{\Delta R}{R_0} \approx -\frac{2\Delta r}{3(r_{om}-r_{im})}$, which allows us to relate  ${\cal R}_e$ to  ${\cal R}_m$   as ${\cal R}_e \approx \frac{2r_{im}}{3(r_{om}-r_{im})}{\cal R}_m$. Both experimental responsivities are of the same order of magnitude  ($\sim 10^{-6}$ Pa$^{-1}$), as listed in Table \ref{table1}, confirming our analysis. This agreement gives us confidence when converting electrical responses to forces and displacements in the experiments. We note, as above, that a linear approximation is an oversimplification, especially for high displacements. Also as above, uniform forces applied  on the shell from the outside result in slightly different electrical responses  compared  to forces on the cage applied  from the inside   due to the complex deformation patterns of the shell (see ESI). In what follows, we take this into account and use the appropriate calibration when we present results involving forces exerted simultaneously both from the inside and outside of the seeding well (e.g., Fig. \ref{fig:fig5}). 

\begin{table*}
\caption{\label{tab_devices} Mechanical and electromechanical characteristics  of the measured devices.}
\begin{ruledtabular}
\begin{tabular}{cccccccccccc}
Device  & $k_{eff}$  & ${\cal R}_m$ &  $\varepsilon_{max}$ &  $\dot \varepsilon_{max}$ & ${\cal R}_e$ & $R_0$ & $\Delta R_{min}$ & $\Delta p_{min}$  & $ F_{min}$ & $\Delta r_{min}$  \\
   & (N/m) &    ($\rm Pa^{-1}$)  & & ($s^{-1}$) & ($\rm Pa^{-1}$) & ($\rm k\Omega$) & ($\rm \Omega /Hz^{1/2}$) & $(\rm Pa/Hz^{1/2})$  & ($\rm \mu N/ Hz^{1/2}$) & ($\rm nm/Hz^{1/2}$) \\
\hline
$20~ \mu \rm m$  planar &  $16.9$ &  $-5.5 \times 10^{-6}$  & $0.15$  & $1.5$ & $-6.86 \times 10^{-6}$ & $43.6$ & $0.67$ & $8.8$ & $0.36$ & $13$  \\
$30~\mu \rm m$ curved &  $31.1$ &  $-4.7 \times 10^{-6}$  & $0.25$  & $2.5$ & $-3.16 \times 10^{-6}$ & $68.9$ & $1.3$ & $12.8$ & $0.52$ & $25$ 
\\
\label{table1}
\end{tabular}
\end{ruledtabular}
\end{table*}

All the relevant experimental and numerical device parameters and response characteristics are presented in Table \ref{table1}. Both the mechanical and electrical responsivities and calibrations are based on linear approximations taking $\Delta r$ as the relevant parameter. We reemphasize that the complex geometry of the device and its deformations need to be taken into account for obtaining accurate predictions of  ${\cal R}_e$,  ${\cal R}_m$  and $k_{eff}$. The very large maximum strain values, $\varepsilon \sim 15-25 \%$, reported here are due to the fact that PDMS shells can withstand  large pressures, $\Delta p \lesssim 800$ mbar. At the highest $\Delta p$ values, the seeding well stretches such that $r_{im} \approx r_{om}$, and the linear approximation described above  is not valid in this regime (see Movie S2).  The bandwidth of  actuation  (i.e., actuation speed) is determined from rise time measurements (e.g., see Fig. \ref{fig:fig5}A below) to be $\approx 10~\rm Hz$, probably limited by the viscoelastic response of the structure. This provides a maximum strain rate of $\approx 2.5/$s. Our electrical measurement setup allows for  a noise floor (sensitivity) for measurement of resistance of  $(\Delta R/R)_{\rm RMS} \approx ~6 \times 10^{-5}$ at a  resistance of $R_0 \approx 50~\rm k\Omega$. This translates into a $2.5~\rm \Omega$ resolution for a 15 Hz bandwidth, which corresponds to a displacement detection limit of  $\Delta r \approx 50 ~ \rm nm$. Besides the practical advantages, this resolution makes resistive monitoring potentially more sensitive than conventional optical monitoring techniques.

\begin{figure*}[t!]
 	\centering
\includegraphics[width=\linewidth]{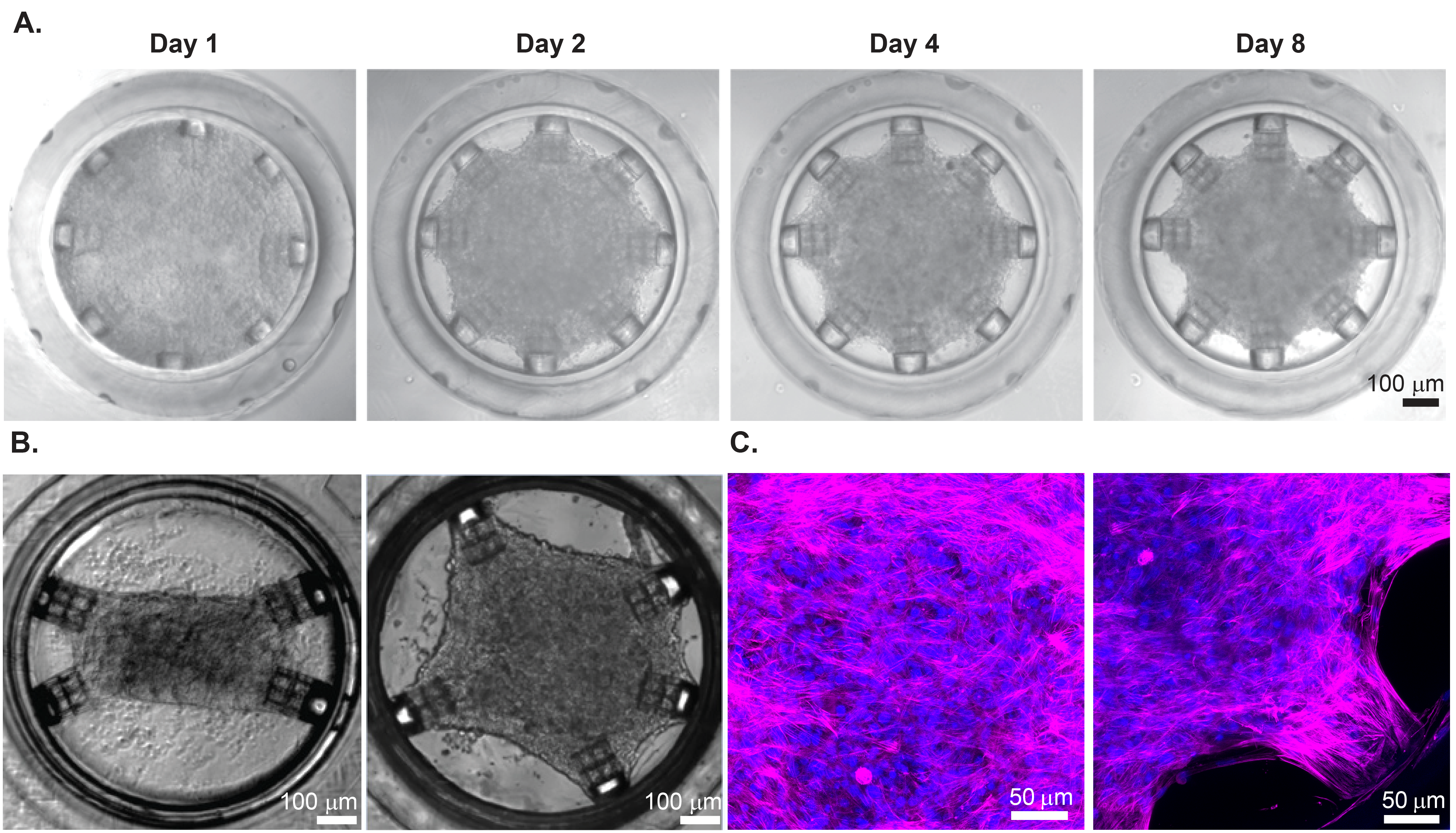}
 	\caption{Tissue remodeling via attachment sites. (A) Microtissue compaction in the device over the course of 8 days. (B) Attachment sites provide cues during tissue remodeling that define the geometry of the freestanding cardiac tissue: rectangular (left), pentagonal (right). (C) Octagonal microtissues were stained with DAPI (blue) to show nuclei and FITC (magenta) to show sarcomeric actinin. Fluorescent imaging confirmed that the  center of the microtissues were suspended within the devices  (left) and their edges made good contact with the cages (right). Scale bars are 100 $\mu$m (A, B) and 50 $\mu$m (C).}
 	\label{fig:fig3}
 \end{figure*}

\subsection{Experiments on Cardiac Microtissues}
 
To demonstrate the use of this platform in  microtissue characterization, we have performed several experiments on hiPSC-derived cardiac microtissues under different mechanical conditioning. To culture the cardiac microtissues, the platform was sterilized and cell-laden hydrogel solution was pipetted into the large media reservoir at the center of the device. After the cells were centrifuged into the four seeding wells, microtissues were allowed to compact over several days. Similar to  the observations of Legant \textit{et al.}\cite{Legant2009} and Boudou \textit{et al.}\cite{Boudou2012}, the cages on the device walls constrained the remodeling of the collagen and acted as attachment points for the cells. Fig. \ref{fig:fig3}A shows the process of tissue compaction in a circular actuator with eight attachment points. Since attachment sites provide physical cues to the cardiac microtissue during remodeling, they play a critical role for defining the geometry of the tissue. This can be seen in Fig. \ref{fig:fig3}B, where tissue compaction around cages with different orientations resulted in rectangular and pentagonal tissues.

To confirm that cells were attached to cages and suspended across the device, microtissues were stained using DAPI and FITC, and imaged using confocal microscopy (Fig. \ref{fig:fig3}C). These images show that sarcomeric actinin wraps around the attachment sites and uses these sites as boundaries during the compaction process. The fluorescent signal  comes only from the image plane, which is focused on the cages. Since cage diameter is 100 $\mu$m, the thickness of the resulting microtissue is $\lesssim 100~\mu$m,  which is expected to have a few layers of cells. The sarcomeric actinin structure suggests that the fibers are  aligned to the 3D printed attachment sites and in between adjacent sites (Fig. \ref{fig:fig3}C, right). However, the center of the tissue appears to lack anisotropic muscle alignment (Fig. \ref{fig:fig3}C, left) possibly due to the isotropic nature of the octagonal design. On day 4-5 after seeding, synchronized spontaneous microtissue contractions were observed for these octagonal, rectangular and pentagonal tissues (see Movie S3). In the rest of the study, we only focus on the octagonal configuration. 

\begin{figure*}[t!]
	\centering
	\includegraphics[width=\linewidth]{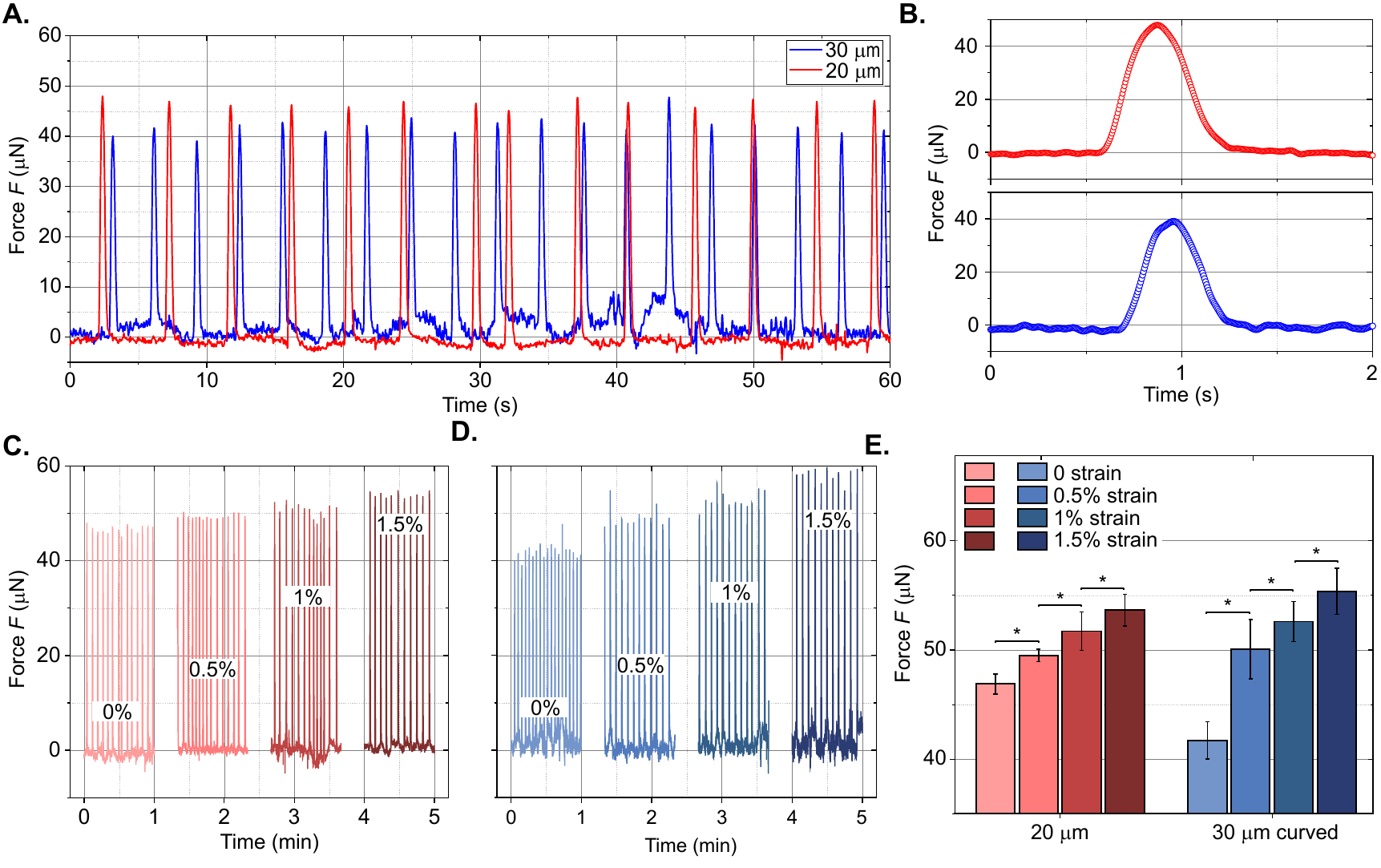}
	\caption[Spontaneous contractions of cardiac microtissues under static strain conditioning]{Spontaneous contractions of  cardiac microtissues under static strain conditioning. (A) Spontaneous contractions  of cardiac microtissues as a function of time in 20-$\mu$m planar (red) and 30-$\mu$m curved (blue) devices for 60 seconds. (B) Representative signals of individual spontaneous beatings from these devices showing muscle contraction (systole) and relaxation (diastole). (C) Spontaneous contractions of a microtissue at different applied strain $\varepsilon$ values between 0-1.5$\%$ in the in 20-$\mu$m planar device. (D) Spontaneous contractions of a microtissue at different applied  $\varepsilon$  between 0-1.5$\%$ in the in 30-$\mu$m curved device. (E) Force clamping data from the two devices, showing that spontaneous contraction force increases with increasing strain. $*$ indicates $p$-value less than 0.0001. The number of beats $n$ analyzed is $n\geq23$ for each strain level.}
	\label{fig:fig4}
\end{figure*}

Fig. \ref{fig:fig4}A displays the contractile force of two octagonal microtissues in the same platform but different devices, as a function of time (also see Movie S4). The force is sensed by the electrical sensor shown in Fig. \ref{fig:fig2}D and converted into Newtons using the above-described calibration. In Fig. \ref{fig:fig4}B, we show typical individual peaks for these microtissues, which contain muscle contraction (systole) and relaxation (diastole) phases of a heartbeat.

Next, we performed a series of force clamping measurements. Figure \ref{fig:fig4}C and \ref{fig:fig4}D  show the microtissue contraction as a function of time under imposed strains in the 20-$\mu$m and 30-$\mu$m  devices, respectively. Spontaneous contractions were monitored for 2 minutes (at least 20 contractions per stretch step) as a constant negative pressure was applied to the transducer to passively stretch the tissue. Although the applied strain levels ($\epsilon \approx 0-1.5 \%$) were relatively low compared to the other studies \cite{Ruan2016,goldfracht2020generating}, the average contraction force data in Fig. \ref{fig:fig4}E  show a clear trend of increase with stretch, consistent with the Frank-Starling law. 

\begin{figure*}[bt!]
	\centering
	\includegraphics[width=\linewidth]{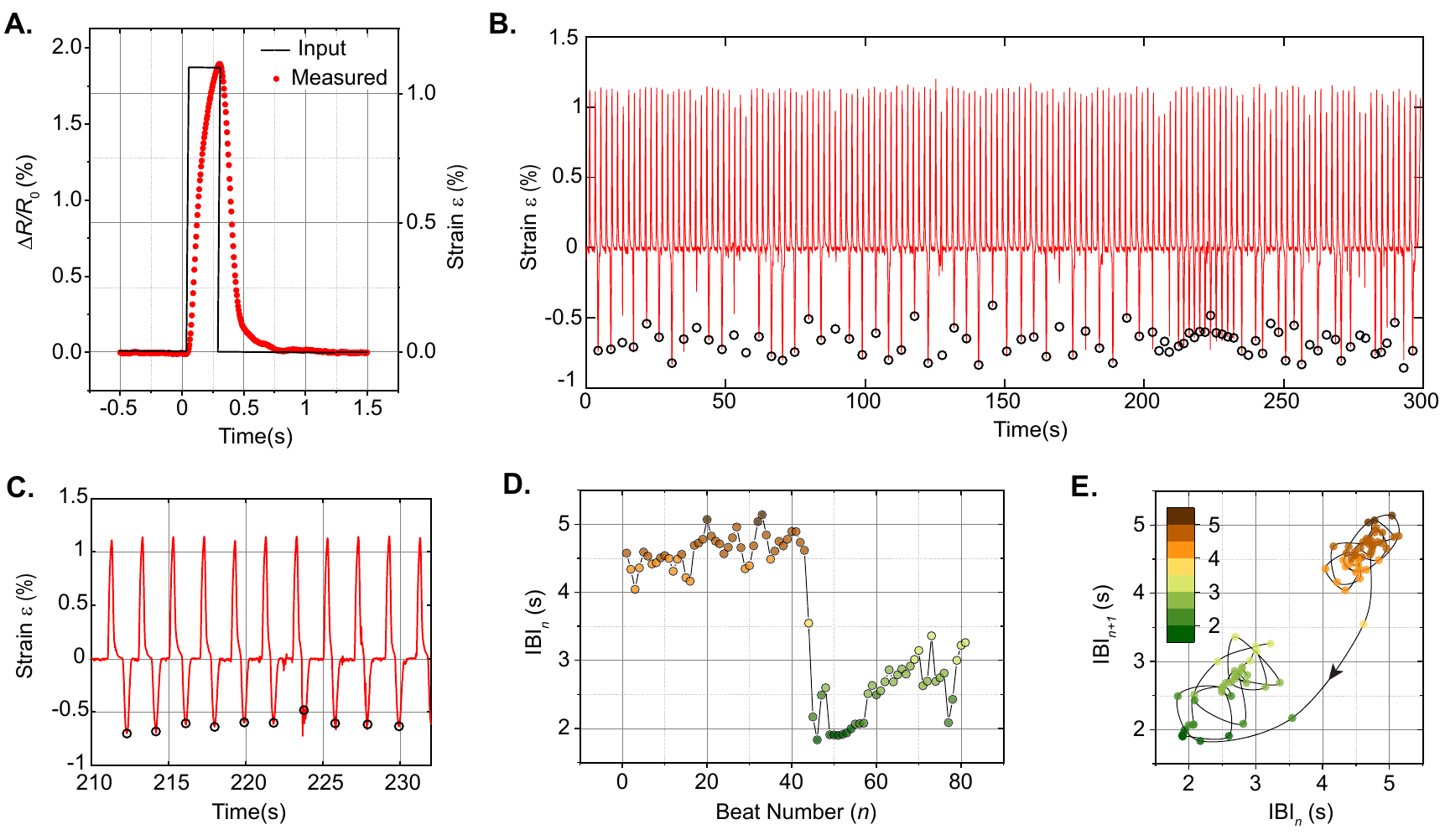}
	\caption[Spontaneous contractions of cardiac microtissues under cyclic strain]{Spontaneous contractions of a cardiac microtissue under cyclic strain in a 20-$\mu$m planar device. (A) The applied  waveform (black line, $\Delta p=-20$ mbar  at 0.5 Hz at a duty cycle of 12.5$\%$) generates a cyclic strain of approximately 1.1$\%$. The response of  the device (red symbols) is detected electrically (left $y$ axis)  and converted  to strain (right $y$ axis) using the calibration procedure. (B) A 5 min data trace showing the electrical response coming from the device with the cardiac microtissue, while pacing at 0.5 Hz.  The large positive peaks correspond to the  external  oscillatory strain. The spontaneous contractions of the microtissue result in the negative peaks.   (C) A close up of the region in (B) in which frequency locking is observed. (D) Interbeat intervals $\rm IBI_{n}$ found from the data in (B). (E) Poincaré map generated from consecutive interbeat intervals.}
	\label{fig:fig5}
\end{figure*}

We now turn to a study of mechanical pacing in our platform. In this experiment, an external oscillatory strain at a frequency of 0.5 Hz and a maximum strain amplitude of $\approx{ 1.1\%}$ was imposed on the device by modulating the applied pressure, while cardiac microtissues were spontaneously contracting (see Movie S5). The external pacing frequency was selected to be higher than the spontaneous contraction rate in order to see if the cells would adjust their beating frequency to keep up with the external perturbation. The applied pressure was a rectangular wave at 0.5 Hz with an $12.5\%$ duty cycle at an  amplitude of $\Delta p=-20~\rm mbar$. Fig.\ref{fig:fig5}A shows the input to the pump to generate the pressure waveform  (black line) and the measured electrical response in the microfluidic device  (red circles, left $y$ axis), with the corresponding strain values indicated on the right $y$ axis. The bandwidth limitation resulting in the distortion of the rectangular wave is primarily due to the viscoelastic response as discussed in ESI. Figure \ref{fig:fig5}B is the output of the resistive sensor over 300 seconds of applied pacing. Our calibration procedure allows us to covert the data into  units of strain: the input pulses increase the strain on the microtissue up to $\epsilon_{r}\approx 1.1 \%$, and the active contractions of the tissue appear as compressive strain pulses of $\epsilon_{r} \approx -0.75 \%$. We have used a standard peak detection algorithm to detect the peaks  (black circles). In some instances (e.g., at $t= 57, 89, 127, 169$ s), tissue contraction and the strain pulse overlapped in time, resulting in a superposition of the two signals and appearing like a single pulse of $\epsilon_{r}\approx0.9\%$. These peaks were identified and quantified by inspection. Around $t=200$ seconds,  the cardiac microtissue abruptly started to beat faster, with the beats occurring right after external perturbations, as shown in the zoomed in trace in Fig. \ref{fig:fig5}C. Here, we  analyze the rhythm of the peaks instead of the contractile amplitude. Following previous studies \cite{wilders1996model,arai2020dynamic,yasuda2020dominant}, interbeat intervals (IBI) are generated as $\rm{IBI_{n}}=t_{n}-t_{n-1}$, where $t_n$ is the moment in time the beat occurs. Fig. \ref{fig:fig5}D displays the IBIs as a function of beat number $n$ over the course of 5 minutes, which is determined from the data in Fig. \ref{fig:fig5}B. After about 40 beats, there is a clear transition between IBIs, from around 5 seconds to 2 seconds. For a short while,   the cells are able to keep up with the applied mechanical pacing. The Poincaré map in Fig. \ref{fig:fig5}E  shows  the same transition phenomenon more clearly. At early times in this data trace, the average frequency of the spontaneous beating was around 0.2  Hz, and the mechanical pacing frequency did not seem to effect the rhythm of the cardiac microtissue. The dramatic change to an average frequency of 0.5 Hz occurred over three cycles. The high-frequency synchnronized beating of the microtissue appears to be more noisy as compared to its spontaneous beating. The noise in the beating frequency may be a sign of an immature microtissue and requires further investigation \cite{cohen2020long}.

\begin{figure*}[t]
     \centering
     \includegraphics[width=\linewidth]{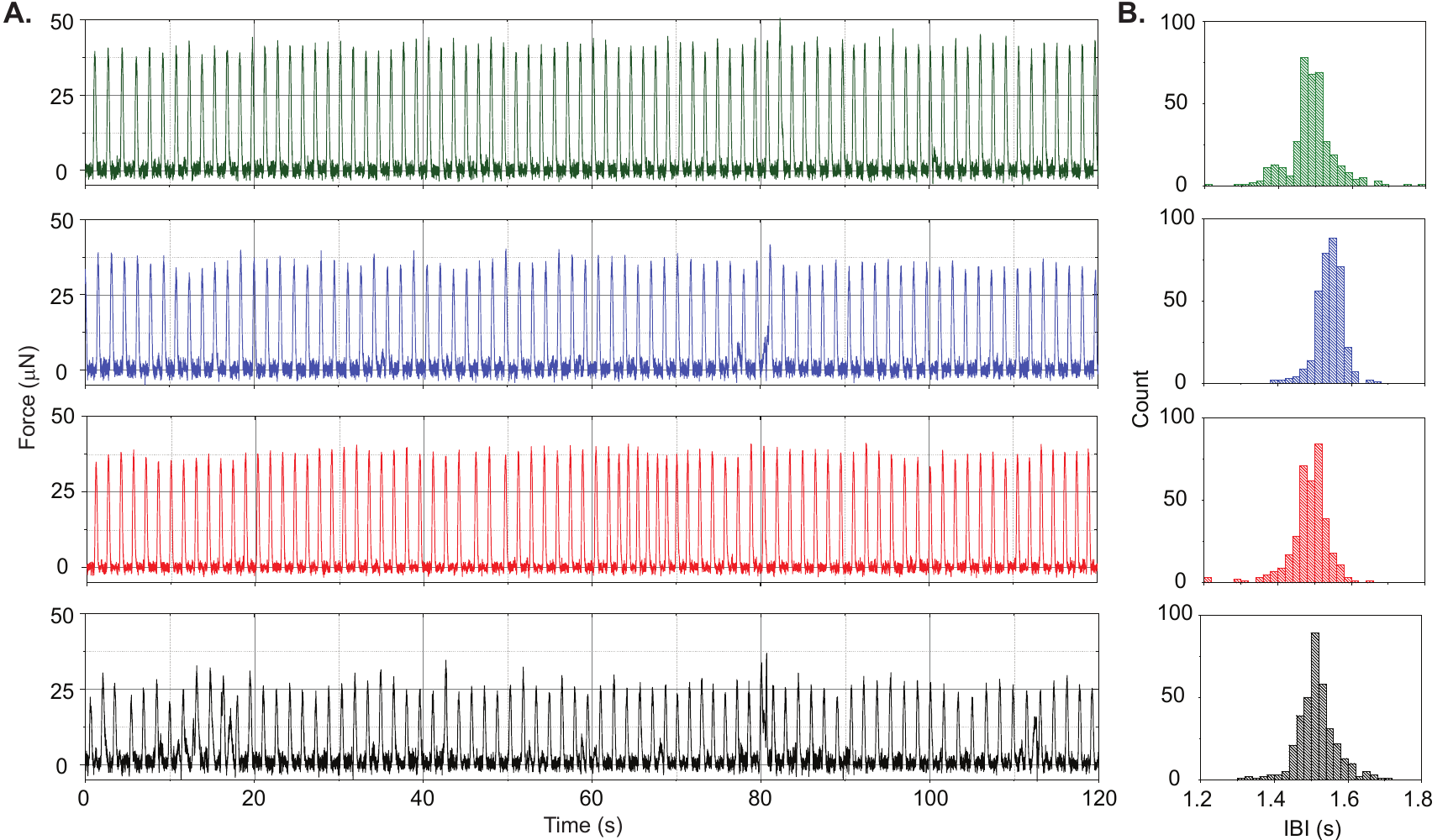}
     \caption{Parallel sensing of active contractile forces from all four devices on a single platform. (A) The signals from four  custom  analog lock-in amplifiers  operating at slightly different reference frequencies (black: 30-$\mu$m-thick curved, 220 Hz; red: 20-$\mu$m-thick, 260 Hz; blue: 20-$\mu$m-thick, 320 Hz; olive: 30-$\mu$m-thick curved, 290 Hz). (B) Histogram of interbeat intervals (IBI) from 550 seconds of data (black: $\rm IBI=1.51 \pm 0.07~s$, red: $\rm IBI=1.47 \pm 0.09~s$, blue: $\rm IBI=1.54 \pm 0.05~s$, olive: $\rm IBI=1.49 \pm 0.07~s$) with the number of analyzed beats $n>360$ for each histogram.}
     \label{fig:fig6}
 \end{figure*}

In Fig. \ref{fig:fig6}, we demonstrate  simultaneous electronic detection of active contractions coming from all four devices on a single platform. In order to accomplish this, we used four home-made portable lock-in amplifiers that are tuned to slightly different reference frequencies in order to avoid electrical cross-talk  between the devices (see ESI). To  convert the electrical signal to force, we used the above-described  approach, first simulating $k_{eff}$ for each device in FEM,  then determining  $\Delta r$ from optical measurements (Movie S6), and finally measuring the  linear electrical calibration curve, $\Delta R \propto -\Delta r$. In summary, the above-mentioned steps allowed us to determine the force amplitude of the contractile peaks coming from each microtissue consistently.  In Fig. \ref{fig:fig6}B, we show the histograms of the interbeat intervals. Cardiac microtissues grown on this platform had an average beat rate of $0.66 \pm 0.01 ~\rm Hz$, while each cardiac microtissue displayed a slightly different  characteristic rhythm.

\section{Discussion}

We begin our discussion by highlighting the several novel aspects of our fabrication approach and device  \cite{Lamont2019,lin2018soft,Marino2018}. The use of DLW lithography to fabricate the negative master molds has  allowed  us to create unique  features that would  be very difficult and even impossible to fabricate using standard lithography. The first of these features is the curved shells (Fig. \ref{fig:fig1}C). The curved shells can be made thicker compared to planar shells while still providing good mechanical responsivity. Thicker shells are more robust and easier to bond to the glass substrate. Control over the heights of   different device regions on the mold has enabled us to fabricate the entire  platform after a single-step molding process. Achieving this  feature is difficult with standard microfabrication, since that would require development of a multi-layer fabrication procedure with precise alignment in between steps. DLW has also allowed us to create functional microstructures on the sidewalls of the device.  DLW of attachment sites at precise locations on the sidewalls of the PDMS shell has provided the mechanical stiffness needed for 3D cardiac tissue remodeling and its subsequent mechanical stimulation. Control in the angular degree of freedom $\theta$ during DLW  has further enabled  geometric  control of the morphology and shape of the cardiac microtissues, as shown in Fig.\ref{fig:fig3} and Movie S3.    
 
Performance metrics of the device are shown in Table \ref{table1}. The spring constants $k_{eff}$ of the shells are about one order of magnitude larger than tall PDMS cantilever gauges or suspended wires \cite{zhao2019platform}, and are of the same order of magnitude as shorter micropost arrays \cite{Hinson2015}. Maximum achievable strain here ($\epsilon_{max}\approx 15-25\%$) is comparable with other  3D cardiac stretchers based on pneumatic actuators ($\epsilon_{max}\approx 15\%$) \cite{marsano2016beating,Parsa2017}. However, transduction of applied pressure to strain in our platform is more  linear compared to other reported pneumatic 3D  platforms,  making calibration  straightforward. The strain rates achievable here are comparable to those  reported in electromagnetic cell stretchers and other hydraulic/pneumatic platforms but  are lower than 2D cell-stretchers based on dielectric actuators \cite{poulin2018ultra,Imboden2019}. 

Monitoring the contractile forces of a cardiac microtissue electrically by  microfluidic strain gauges has some advantages over optical monitoring. For optical monitoring, one typically needs a microscope equipped with a $\rm CO_{2}$ and temperature controlled environmental chamber. Electrical sensing can be accomplished by a simple Ohmmeter and will allow real-time detection and long-term monitoring of  contractile forces inside any regular cell culture incubator. Optical detection is limited by the spatial resolution  and the field of view of the objective. Electrical monitoring is scalable in that signals from many devices distributed over a large area can  be detected, as shown in this work, allowing for high-throughput contractility assays. In the near future, it will be possible to further optimize the platform for  direct administration of drugs  while electrically monitoring contractions of a large array of microtissues.

The Frank-Starling effect describes the relationship between tissue length and contraction strength and is commonly used as an indicator of healthy heart tissue \cite{zhang2013tissue, huebsch2016miniaturized,Ribeiro2019,Schwinger1994,Helmes2016}. This effect has been observed in hiPSC-derived cardiac tissues in conjunction with electrical stimulation \cite{Mannhardt2016,Uesugi2019,Tulloch2011,Ruan2016,goldfracht2020generating}. Here, our smaller microtissue constructs have displayed similar Starling curves even for miniscule increases in tissue length ($\epsilon_{r} \approx 0.5\%$, $\Delta {r} \approx 4 ~\rm \mu m$) and in the absence of electrical pacing.  

From the start of active contractions,  maturation and remodeling of cardiac tissue is believed to be strongly influenced by the cyclic mechanical stress in the tissue \cite{kofron2017vitro}. Previous research has shown that application of cyclic strain on 3D engineered cardiac microtissues over a time period of days to weeks promotes hypertrophy \cite{Parsa2017}, electrical and mechanical coupling between the cardiomyocytes in the tissue, as well as stem-cell differentiation\cite{kofron2017vitro,marsano2016beating}. However, the subtle aspects of synchronization between  cardiac microtissue contractions and external periodic  stimuli  has not yet been investigated. Recent work has shown that individual neonatal rat cardiomyocytes can synchronize with periodical  strains and beat at the frequency of the external strain. \cite{Nitsan2016}.  This study on entertainment of individual rat cells has not yet been extended to freestanding cardiac microtissues. In order to achieve  synchronized beating, it is believed that mechanical perturbations on the tissue must closely mimic the \emph{in vivo} deformations generated by neighbouring cells \cite{chiou2016mechanical}. Ventricular tissue undergoes cyclic strain during each pressure-volume cycle \emph{in vivo}. Our experiments shown in Fig. \ref{fig:fig5} aim to mimic this cyclic strain. Although preliminary, we have observed periods of synchronization between the tissue contractions and the oscillatory mechanical perturbations. 

\section {Conclusion and Outlook}

In this work, we have introduced a versatile 3D heart-on-a-chip platform with integrated transducers. The platform and its unique fabrication approach will allow users to tune the mechanical properties of the device, including the  sensitivity and stiffness of the cellular micro-environment. The first route to tuning device properties is to change the linear dimensions. For instance, mechanical responsivity of the platform can be enhanced by simply decreasing the radius of the seeding well $r_{c}$ or making thinner and taller walls. Similarly, electrical responsivity can be increased by decreasing the width of the annular microchannel. The second possibility is to use the almost limitless design and fabrication capabilities  offered by DLW. While  two shell types were investigated here, the responsivity can be tailored by, for example, varying the thickness locally, or printing exotic structures, such as buckled shells. In order to promote sarcomere alignment and uniaxial contraction, the structure and orientation of the attachment sites can easily be modified to a more anisotropic configuration (Fig. \ref{fig:fig3}B). 

This manuscript has primarily focused on  the fabrication and mechanical and electrical characterization of the platform. Additional work is needed to thoroughly assess the quality of the microtissues generated in our platform. In the near future, response to inotropic drugs  will be evaluated to  determine the overall health of the cardiac microtissue. Electrophysiological measurements will be performed to assess action potential duration and conduction velocity. The design  of our platform facilitates these important tissue characterization tasks; drugs can be administered easily into the open seeding wells, and  optical access allows for  fluorescent microscopy in order to investigate action potential generation and calcium dynamics. Simultaneous monitoring of these parameters under applied  strain will enable further studies of cardiac mechano-electric coupling. 

The design of the platform can also be optimized for high-throughput screening applications. A seeding well is $\sim 0.8 ~\rm mm$ in diameter and the footprint of an individual  device is $\sim 1 ~\rm mm$ in diameter. With these dimensions, it will be possible to fabricate   one device per  well of a standard 384 well plate ($d\sim 3.5 ~\rm mm$). Alternatively, a similar device density can be achieved by packing 16 devices as a $4 \times 4$ array for each  well of a standard 24 well plate ($d \sim 16 ~\rm mm$). With the present approach, the bottleneck to fabrication speed will be the 3D printing  of the attachment sites on the sidewalls of each device,  with each attachment site taking $\sim ~\rm 1$ minute to print.

Further device functionality can be achieved  by taking advantage of the purely electrical readout of active contractions.  A closed-loop feedback system can be implemented to adjust mechanical pacing based on tissue response. Stimulation electrodes can  easily be integrated to incorporate electrical pacing in addition to mechanical pacing. Furthermore, electrical and mechanical stimulation at the same pacing frequency but adjustable phase with respect to each other is also possible. Finally, it may be possible to de-couple the actuation and sensing ``ports" by modifying the single cylindrical shell into two symmetric hemicylindrical shells.   

\section {Materials and Methods} 
\subsection{Device fabrication}
A commercial DLW system (Nanoscribe Photonic Professional GT) with a $25 \times$ (NA=0.8) immersion objective is used to print the master mold on  IP-S negative photoresist (Nanoscribe, GmbH) dropcast on a silicon substrate.  To ensure adhesion between the mold and the substrate across the entire footprint of the structure, the silicon substrate is treated with 3(trimethoxysilyl)propyl acrylate (TMPA) prior to drop-casting the photoresist.  The DLW process  starts at a depth of 5 $\mu$m below the surface of the substrate and proceeds upwards in $z$ direction. The $25 \times$ objective enables a maximum printing volume of $l \times w \times h \approx 400 \times 400 \times 300 ~\rm \mu m^3$ (in $x, y$ and $z$, respectively) with a minimum voxel  of diameter $0.6~\rm \mu m$ in the $xy$-plane and height $1.5~ \rm \mu m$ along the $z$-axis. The maximum print regions can further be stitched together to produce even larger structures that can cover a 4-inch wafer. In this work, the mold design for the entire platform (Fig. 1C) measures  $18 \times 10 \times 0.5 ~\rm mm^3$ (in $x, y$ and $z$). After printing, remaining monomeric photoresist is rinsed from the mold by propylene glycol monomethyl ether acetate (PGMEA). 

Once the negative master molds are completed, they are treated with trichloro(1H,2H,2H-perfluoroocytl) silane (TPFOS) overnight to reduce PDMS adhesion. Subsequently PDMS (10:1, Sylgard 184) is cast onto the molds. To ensure that the seeding well region remains open through the thickness of the PDMS, a piece of glass with added weight is placed on top of the uncured PDMS in the DLW mold, resting on the top surface of the 500 $\mu$m tall seeding well negative. PDMS is cured on a hotplate at 150$^{\circ}$C for 15 minutes and then demolded once it  cools down to room temperature. Care must be taken to ensure that the thin PDMS shells do not rupture and stay behind in the mold during the demolding. The actual height of the fabricated shells is found to be $\approx$280 $\mu$m due to some PDMS shrinkage and the  printing offset used to ensure adhesion. 

After demolding, PDMS devices are treated with TMPA to ensure adhesion between the inner sidewalls of the seeding well and the 3D  small cylindrical attachment sites (``cages'') to be printed on the inner sidewalls of the shells. The 10-degree inward taper helps to avoid laser power loss due to shadowing from the surrounding structures and to ensure that the cages have good adhesion to the PDMS. These cages are 100 $\mu$m in diameter and 200 $\mu$m in length and are printed using DLW in pentaerythritol triacrylate (PETA) with 3 wt$\%$ Irgacure 819 (BASF) photoinitiator.  

After the ``cage" printing, microfluidic inlets are punched into the PDMS  by using a biopsy punch ($d=0.75 ~\rm mm$, World Precision Instruments). Following optical inspection, the devices are plasma treated (PDC-32G, Harrick Plasma) with air, 45 seconds under 10.5 W RF power, then bonded to an electrode-patterned glass substrate. The electrodes are patterned by electron beam evaporation: a thin film of Ti/Au (20 nm/ 80 nm) is deposited on a $22 \times 22$ mm$^2$ coverglass through a stainless steel shadow mask. We have found that it is more reliable to bond the thin shells to a cover glass that is spin-coated with a thin layer of  PDMS as compared to  bare glass. Before spin coating the PDMS, the cover glass with  gold electrodes is selectively masked with  scotch tape to ensure that the electrodes are exposed in the appropriate  regions (contact pads and electrode ``tips” closest to the cylindrical shell where sensing occurs). Next, the cover glass spin-coated with PDMS ($\sim$20-$\mu$m-thick film)  is plasma treated, and the tape is removed. Then the PDMS structure is bonded to the glass, and  the sample is baked at 150$^{\circ}$C for 15 minutes. Once the PDMS structure is bonded to  the glass, a thicker PDMS piece with a media reservoir is bonded on top of the device array. Figure \ref{fig:fig1}B (right) shows the final device ready for cell seeding and testing. A 3D printed (Formlabs) holder is used to connect the electrodes to wires.

The fabricated platform is connected to a microfluidic flow control system (OB1-MK3, Elveflow) for actuation.  Fluorinated Ethylene Propylene (FEP, Cole-Parmer) tubing and custom-made stainless steel pipes ($d=0.9 ~\rm mm$, New England Small Tubing) are used for the connections. For the  pressure range used (-900 to 1000 mbar), the flow system provides a pressure stability  of $\sim100$ $\mu$bar and response time of $\sim 10$ ms.  For calibration of electrical signals with respect to applied pressure, a 200 Hz sampling rate is used for both pressure regulation and electrical sensing.

\subsection{Computational Models}

Finite element models of the devices are carried out in  COMSOL Multiphysics Platform (version 5.5, COMSOL, Inc). Young's moduli of PDMS and PETA are taken as $2.2$ MPa \cite{Johnston2014} and $260$ MPa \cite{jayne2018}, respectively; Poisson's ratio of PDMS  is taken as 0.48 and of PETA as 0.40. To mimic adhesion between PDMS and the bottom glass surface, we assume a fixed boundary at the bottom surface of the model and  keep the rest of the boundaries in the model free. To mimic  vacuum, we apply an  outward normal stress $\Delta p$ on the outer walls of the PDMS shells ranging from -50 to 50 mbar in 10 mbar increments. To mimic contractile forces exerted by the tissue, we apply a normal outward stress on each of the circular surfaces of cylindrical PETA cages, ranging from 0-50 mbar in 10 mbar increments. We then convert the stress to  force using $F=\sigma \pi r_{cage}^2$, with $r=50~\rm \mu m$  being the radius of a PETA cylindrical cage. Details of the computational simulations are provided in ESI.       

\subsection{Electrical Measurements}
Electrodes are connected to external electronics by using a 3D printed holder for the platform and pressure connect pins. A lock-in amplifier (SR830, Stanford Research Systems) is used for the electrical measurements. An equivalent circuit diagram is shown in ESI. Here, the lock-in reference oscillator output is set to a frequency of $f_c \approx25$ Hz and an amplitude of 5 V. The output is connected to a 10-M$\Omega$ resistor to create a current source of  amplitude  $I \approx500$ nA. This current is passed through the microchannel to ground. The second pair of integrated electrodes are connected to the lock-in input (in parallel with the 10-M$\Omega$ input resistance), and are used to sense the voltage drop across the sensing region. The resistance of the channel can be expressed as $R(t)=R_0+ \Delta R(t)$, where $R_0$ is the initial time independent resistance and the time-dependent $\Delta R(t)$ comes from the PDMS shell deflections due to external actuation or cell beating in the sensing region of the device. The lock-in amplifier, after mixing the input voltage down with the oscillator reference frequency, outputs the voltage $V(t) \approx IR_0+I \Delta R(t)$. See ESI for the calculation of $R(t)$ from $V(t)$. The initial voltage can be approximated as $I R_0$, and the voltage fluctuations are $\propto I \Delta R(t)$ at frequencies within $0 \leq f \leq $ BW. \cite{Kara2018}. Here, BW is the bandwidth set by the time constant of the lock-in amplifier, and BW $\approx$ 15 Hz.  

\subsection{Image Collection}
Images and videos for calibration are obtained in an  inverted microscope (Axio Observer, Carl Zeiss) using a $20\times$ objective, an AxioCam 503 mono camera (Carl Zeiss), and ZEN image acquisition software (Carl Zeiss). Microtissue beating videos are acquired with a $10\times$ objective on a Nikon Eclipse Ti (Nikon Instruments, Inc.) microscope which is equipped with a live cell incubator. Tissue fluorescence images were acquired using the confocal multiphoton microscope Leica TCS SP8 MP, operated in single photon mode, using a 25x and a 40x water immersion lens.

\subsection{Data Acquisition and Processing}
The output signals from the lock-in amplifier are recorded using a data acquisition card (NI 6221, National Instruments) through a LabVIEW (National Instruments) Virtual Instrument interface. The sampling rate for data collection is 200 Hz for the measurements on individual devices and 100 Hz (for each device) for the simultaneous measurement of 4 devices on the platform. The experimental data are analyzed using OriginPro 2018 (MicroCal Software) and MATLAB (MathWorks). When needed, the lock-in output is high-pass filtered above 0.1 Hz in order to remove the low-frequency drifts; high frequency noise is digitally smoothed by FFT low pass filtering above 20 Hz or Savitzky-Golay averaging (n=10, second order) for peak analysis, without significantly effecting the peak amplitude. After pre-processing, peak detection is performed by detecting local maxima above a threshold ($>40\%$) based on taking the first derivative of the signal. For the baseline subtraction in simultaneous monitoring experiments, an asymmetric least squares smoothing method module of OriginPro is utilized (asymmetric factor = 0.001, threshold $\rm <0.05$, smoothing factor $\rm >4$, number of iterations = 10).    

\subsection{hiPSC Culture and Cardiomyocyte Differentiation}
The human induced pluripotent stem cells (hiPSCs) are created from the PGP1 donor from the Personal Genome Project, kindly shared by the Seidman Lab at Harvard Medical School. Wild type human induced pluripotent stem cells (hiPSC) were maintained in complete mTeSR1 medium (Stem Cell) and differentiated to the cardiomyocyte lineage in RPMI 1640 medium (Gibco) supplemented with B27 minus insulin (ThermoFisher) by sequential targeting of the WNT pathway - activating WNT pathway using 12 $\mu$M of CHIR99021 (Tocris) in Day1 and inhibiting WNT pathway using 5 $\mu$M of IWP4 (Tocris) in Day3 and Day4. Cardiomyocytes were isolated after showing spontaneous beating (usually between Day9 to Day14) using metabolic selection by adding 4 mM of DL-lactate (Sigma) in glucose free RPMI 1640 medium (Gibco) for four days. Following selection, cardiomyocytes were maintained and assayed in RPMI 1640 medium supplemented with B27 (ThermoFisher) and used to make cardiac tissues between 20 to 30 days post initiation of differentiation.

\subsection{Cardiac Tissue Generation}
The PDMS devices were plasma treated (EMS 1050X, EMS Quorum) with ambient air, between 30-60 seconds at 100W and 0.6 mbar. Chips were then sterilized in 70$\%$ ethanol for one hour followed by washing in sterile deionized water for 30 minutes.  The sterilized chips were then treated with 2$\%$ pluronic F127 for 30 minutes at room temperature to prevent cell-laden hydrogel adhesion to PDMS. Cardiomyocytes were dissociated after trypsin digestion and mixed with stromal cells (human mesenchymal stem cells, hMSCs) to enable tissue compaction. A suspension of 1 million cells (90$\%$ cardiomyocytes and 10$\%$ hMSCs) within 2.25mg/mL liquid neutralized collagen I (BD Biosciences) was added per chip  on ice, then centrifuged to drive the cells into the micropatterned tissue wells. Excess collagen and cells were removed by aspiration before incubating at 37 $^{\circ}$C to induce collagen polymerization. The tissue culture media consist of DMEM (Corning) with 10$\%$ Fetal Bovine Serum (Sigma), 1$\%$ GlutaMax (Gibco), 1$\%$ Non-Essential Amino Acids (Gibco) and 1$\%$ Penicillin-Streptomycin (Gibco) was then added to the seeding well. Cells compacted the collagen gel over several days and testing was performed 5 days post seeding.

\subsection{Immunostaining}

Tissues were fixed right after imaging on day 5 using a 4$\%$ paraformaldehyde (PFA) solution in phosphate buffered saline (PBS) for 15 minutes. The PFA solution was removed and the tissues were washed with PBS 3 times. The cell membranes were permeabilized using a PBS solution with 2$\%$ bovine serum albumin (BSA) and 0.3$\%$ Triton-X for 15min at room temperature, followed by 3 PBS washes and 1h in PBS with 2$\%$ BSA at room temperatures. The tissues were washed with PBS 3 times and were stored overnight in a PBS solution with 1$\%$ BSA and the primary antibody for sarcomeric $\alpha$-actinin (ab137346, abcam) at 4ºC, followed by 3 PBS washes and overnight staining at 4ºC in PBS and 1$\%$ BSA with DAPI for nuclei, phalloidin for actin and the secondary antibody for $\alpha$-actinin. The tissues were then washed with PBS 3 times and stored in PBS at 4$^{\circ}$C until imaging.

\subsection{Experimental Setup for Cell Monitoring and Pacing}
During cell seeding right before centrifugation, 1x PBS was added to the actuation and electrode channels in the PDMS device. A small PBS reservoir was included in device design (separate from the media reservoir) to ensure these channels remained hydrated until testing. Once cell-laden devices were ready for testing, they were removed from an incubator and positioned onto a custom sample holder equipped with spring-loaded pins that can be pushed against the gold-patterned contact pads of the device being tested. This sample holder was then placed in a humidity and CO$_{2}$ controlled chamber where the microfluidic pump was connected to the device inlet. At this stage, microtissue contractions within the device could be passively monitored, or pressure differentials could be applied to actively stretch/compress the microtissue while simultaneously monitoring the active contractions.

 \section*{Author contributions}

R.K.J., M.Ç.K., and K.Z. contributed equally to this study. R.K.J., M.Ç.K., K.Z., C.S.C., K.L.E. and A.E.W. conceived the project. R.K.J and M.Ç.K. designed and fabricated the integrated platform. K.Z. differentiated human iPS cells and generated the cardiac microtissues. M.Ç.K. and R.K.J. developed the finite element model. M.Ç.K., N.P., R.K.J., D.J.B, A.E.W. and K.L.E devised the electrical measurements. M.Ç.K., R.K.J., K.Z. and C.M. conducted the experiments and collected data. M.Ç.K., R.K.J., K.Z., C.M., A.E.W and K.L.E. interpreted data and wrote the manuscript. 

\section*{Conflicts of interest}
There are no conflicts to declare.

\section*{Acknowledgements}
This work was supported by NSF CELL-MET ERC award no. 1647837. We acknowledge the support from BU Photonics Center. We thank Pablo Del Corro for his help in building the portable lock-in amplifiers. We thank Seidman Lab for kindly sharing their hiPSCs cell lines. 

\pagebreak




\bibliography{ms.bbl} 

\end{document}


\title{Supplementary Material for\\ ``Direct laser writing for cardiac tissue engineering: a microfluidic heart on a chip with integrated transducers''}

\author{R. K. Jayne}
\altaffiliation{These authors contributed equally to this work.}
\affiliation{\BU}
\affiliation{Photonics Center, Boston University, Boston, MA 02215, USA.}
\author{M. Ç. Karakan}
\altaffiliation{These authors contributed equally to this work.}
\affiliation{\BU}
\affiliation{Photonics Center, Boston University, Boston, MA 02215, USA.}
\author{K. Zhang}
\altaffiliation{These authors contributed equally to this work.}
\affiliation{Department of Biomedical Engineering, Boston University, Boston, Massachusetts 02215, USA}
\affiliation{Wyss Institute for Biologically Inspired Engineering, Harvard University, Boston, MA 02115, USA.}
\author{N. Pierce}
\affiliation{Photonics Center, Boston
University, Boston, MA 02215, USA.}
\author{C. Michas}
\affiliation{Department of Biomedical Engineering, Boston University, Boston, Massachusetts 02215, USA}
\affiliation{Photonics Center, Boston
University, Boston, MA 02215, USA.}
\author{D. J. Bishop}
\affiliation{\BU}
\affiliation{Department of Biomedical Engineering, Boston University, Boston, MA 02215, USA.}
\affiliation{Division of Materials Science and Engineering, Boston University, Boston, Massachusetts 02215, USA}
\affiliation{Department of Electrical and Computer Engineering, Boston University, Boston, MA 02215, USA.}
\affiliation{Department of Physics, Boston University, Boston, MA 02215, USA.}
\author{C. S. Chen}
\affiliation{Department of Biomedical Engineering, Boston University, Boston, Massachusetts 02215, USA}
\affiliation{Wyss Institute for Biologically Inspired Engineering, Harvard University, Boston, MA 02115, USA.}
\author{K. L. Ekinci}
\email[Electronic mail:]{ekinci@bu.edu}
\affiliation{\BU}
\affiliation{Photonics Center, Boston
University, Boston, MA 02215, USA.}
\author{A. E. White}
\email[Electronic mail:]{aew1@bu.edu}
\affiliation{\BU}
\affiliation{Department of Biomedical Engineering, Boston University, Boston, MA 02215, USA.}
\affiliation{Division of Materials Science and Engineering, Boston University, Boston, Massachusetts 02215, USA}
\affiliation{Department of Physics, Boston University, Boston, MA 02215, USA.}

\date{\today}
\maketitle

\tableofcontents

\newpage

\section{Finite Element Simulations}

To estimate the relation between the applied pressure and the resulting displacements or the radial strains on the 20-$\mu$m-thick planar and 30-$\mu$m-thick curved PDMS cylindrical shells, finite element models of devices were developed using the Structural Mechanics module of the COMSOL Multiphysics Platform (version 5.5, COMSOL, Inc). The material properties used in the simulations for  PDMS and PETA  are listed in Table \ref{table1} \cite{Johnston2014,Jayne2018}. To model the adhesion between the PDMS and the glass surfaces, we applied a fixed boundary condition to the bottom surface of the structure. We kept the rest of the boundaries free to move, including the top PDMS anchor. The finite element model is meshed using  free tetrahedral elements, with $\sim 2\times 10^6$  elements  generated. The smallest mesh elements are typically  0.3 $\rm \mu m$ in length.

\begin{table}[b]
\caption{\label{table1} Properties of the materials  used in the simulations.}
\begin{ruledtabular}
\begin{tabular}{cccc}
Material & $\rho$ & $E$ & $\nu$ \\
     & ($\rm kg/m^3$) & (MPa) \\
\hline
Polydimethylsiloxane (PDMS) \cite{Johnston2014} & 970 & 2.2 & 0.48 \\
Pentaerythritol triacrylate (PETA) \cite{Jayne2018} & 1190 & 260 & 0.40
\end{tabular}
\end{ruledtabular}
\end{table}

 \begin{figure}[h]
    \centering
    \includegraphics[width=0.8\linewidth]{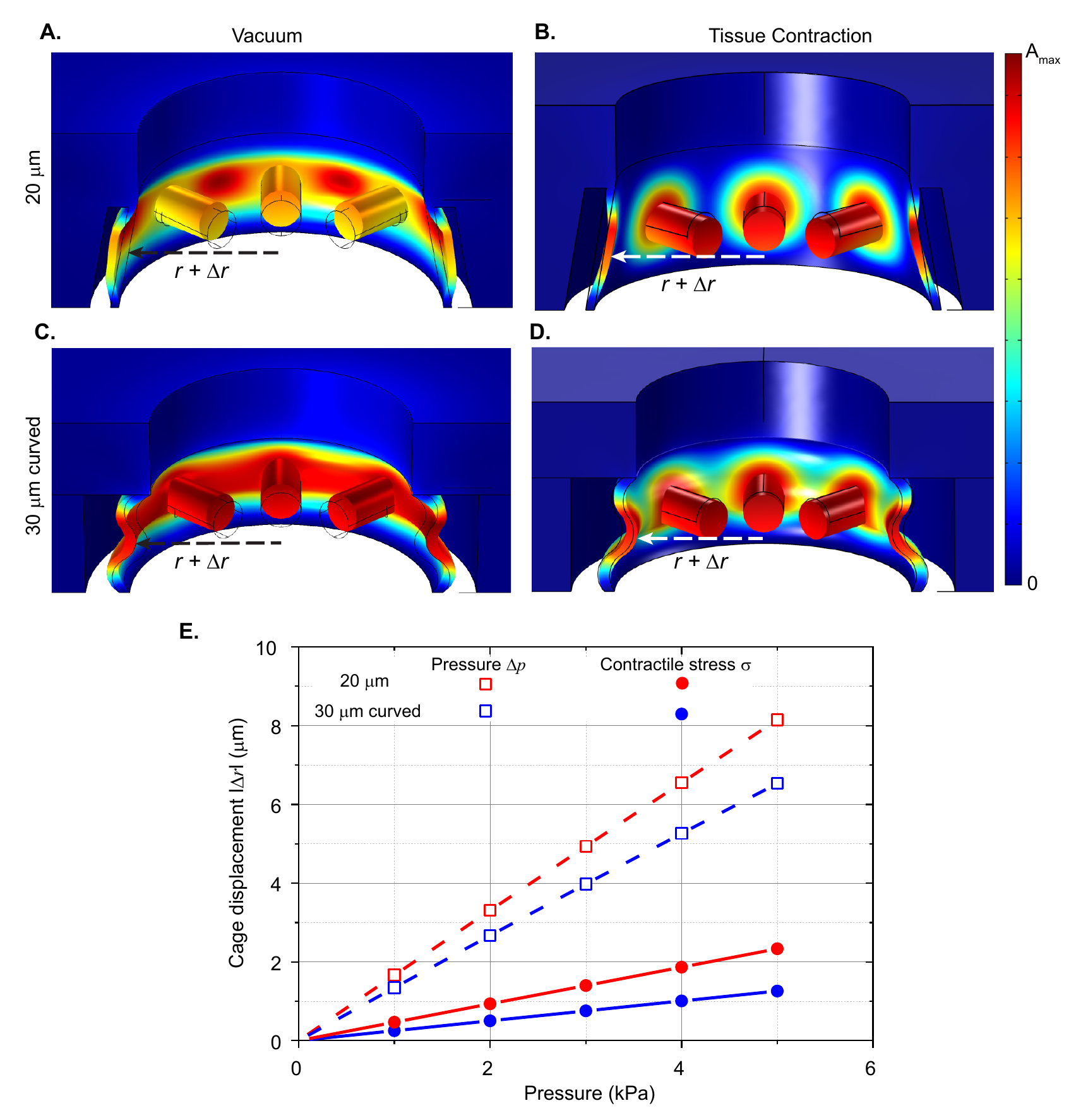}
    \caption{Simulated deformation profile for the 20-$\mu$m-thick planar (A,B) and 30$\mu$m-thick-curved (C,D) PDMS shells under applied vacuum from annular chamber (A,C), and contractile stress $\sigma$ applied from cage surfaces in normal direction (B,D). Cages on the sides are removed after the simulation for clarity (these are shown in Fig. 2B inset). Color represents the normalized displacement amplitude in the radial direction. E: Cage displacement as a function of applied external pressure $\Delta p$ (dashed line) and contractile stress $\sigma = F/(\pi r^2)$.}
    \label{fig:comsol1}
\end{figure}

\subsection{Response to Applied Pressure}

 To estimate the effect of the external pressure differentials on the cylindrical shell, we applied $\Delta p$ as a normal stress upon the outer wall of the PDMS shell.  Fig \ref{fig:comsol1}A shows the  deformation profiles of the 20-$\mu$m-thick planar and 30-$\mu$m-thick curved cylindrical shells under an applied vacuum of $\Delta p=-200 ~\rm mbar$. This $\Delta p$ induced a stretch in the radial direction with $\Delta r> 0$ as shown in Fig. \ref{fig:comsol1}A and C. In these color plots,  only half of each  structure with three attachment sites are shown for clarity. The thin solid lines indicate the initial positions of the walls and the attachment sites, which displace as a result of the deformation of the entire structure.  We have observed that the deformation profile is more uniform across the 30-$\mu$m-thick curved cylindrical shells as compared to the 20-$\mu$m-thick planar ones. Another important point to emphasize is that there is negligible cage deformation compared to the shell deformations. To compare simulations with the calibration experiments,  $\Delta p$ is applied on the  outer wall in the $r$ direction, in 10 mbar increments ranging from -50 to 50 mbar . Subsequently, we extracted the resulting cage displacement $\Delta r$ at the center of the cage ($z=h_{m}/2$) as a function of the applied $\Delta p$.

\subsection{Response to Cardiac Twitches}

Since the engineered cardiac microtissue is anchored by the attachment sites, we assume that the stubs of the cages experience the active contractile forces generated by the  microtissue. Following this assumption, we modeled the contractile force by applying an  outward normal stress on the stubs  of the PETA cages (i.e., the  surfaces of the small cylindrical structures in Fig. \ref{fig:comsol1}A-D) in $-r$ direction. Fig \ref{fig:comsol1}B and Fig \ref{fig:comsol1}D show the  deformation profiles of the 20-$\mu$m-thick planar and 30-$\mu$m-thick curved cylindrical shells under applied tissue forces ($\sigma=40 ~\rm kPa$). We performed simulations for  stress values between 0-10 kPa in 1 kPa increments, which is in the range of  the experimentally observed  values. Then, we extracted the resulting cage displacements at the center of the cage  ($z=h_{m}/2$) as a function of the applied stress. 

The results of all the simulations are shown in Fig. \ref{fig:comsol1}E. The absolute value of the cage displacements $|\Delta r|$ due to the externally applied pressure   $\Delta p$  are shown by open squares whereas those due to normal tissue stresses  $\sigma$  are shown by the filled circles.  The reason that the slopes are different can be understood as follows. $\Delta p$ acts upon the entire outer surface area of the PDMS shell, whereas $\sigma$ acts only upon the surface of the cage stubs. Thus, the overall force  is  more for the case of $\Delta p$ than $\sigma$. The slope difference between the 20-$\mu$m-thick planar and the 30-$\mu$m-thick curved shells is  due to different thicknesses. In order to estimate the effective spring constant $k_{eff}$ experienced by the tissue, we converted the simulated stress to force using $F=\sigma \pi {r_{cage}}^2$ (Fig. \ref{fig:microchannel_cad} and c.f. Fig 2C in main text).

\section{Electrical Measurements}
 
\subsection{Estimation of Electrical Resistance Change as a Function of Cage Displacement}

\begin{figure}
    \centering
    \includegraphics{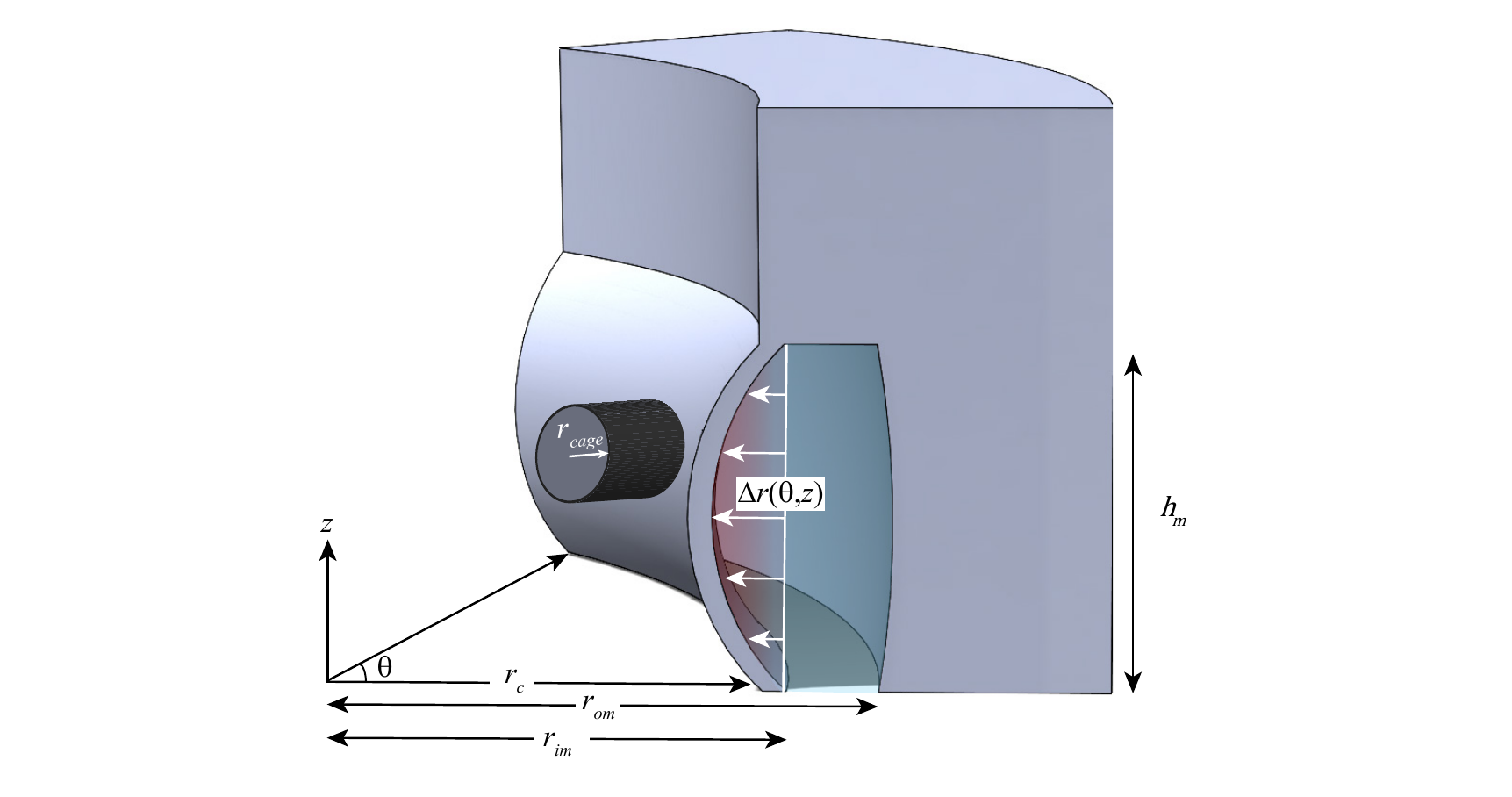}
    \caption{An illustration of a sensing microchannel in cylindrical coordinates. $r_{om}-\ r_{im}$ and $h_{m}$ are the width and height of the sensing microchannel respectively at $\Delta p=0 ~\rm mbar$, and $\theta \approx \pi/4 ~\rm rad$ . Pressure waves and forces exerted by the tissue similarly bends the PDMS shell, changing the microchannel width by $\Delta r (\theta,z)$.}
    \label{fig:microchannel_cad}
\end{figure}

Tissue generated contractile forces and the external pressure both act along the radial axis and   induce bending of the cylindrical shell. Fig. \ref{fig:microchannel_cad} is an illustration of the bending of the shell. Here, width of the sensing microchannel changes by $\Delta r(\theta , z)$. Below, we derive an approximate relation between the average value of $\Delta r$ and the electrical resistance change $\Delta R$. 

First, the deformation $\Delta r(\theta , z)$ will be assumed to be independent of $\theta$ as a simplification. Below, we discuss the validity of this assumption by exploring the deformation patterns in simulations. Since the top and the bottom of the microchannel are fixed, $\Delta r( z)$ should have an approximately parabolic deformation profile in the $z$ direction due to the applied pressure, resulting in an average wall deflection of \cite{ozsun2013non}
\begin{equation}
\overline{\Delta r}  \approx \frac{2}{3}\Delta r\left(z = h_m/2\right) 
\label{eq_1_} 
\end{equation}
Here, $\Delta r\left(z= h_m/2\right)$ is the maximum value of $\Delta r$. Note that the attachment sites are in the middle of the PDMS shell in the $z$ direction (see Fig. 1C(iii-iv) and Fig. 2A in the main text), and $\Delta r\left(z=h_m/2\right)$ is  the displacement that was measured optically in the experiments. 

The angle subtended by the electrical sensing  microchannel is approximately $\pi/4$ rad. The length of the microchannel can be approximated as $L\approx \frac{\pi \left(r_{om}+r_{im}\right)}{8}$, where $r_{im}$ and $r_{om}$ are respectively the inner and outer radii of the microchannel as shown in Fig. \ref{fig:microchannel_cad}. 

Without any perturbation, the electrical resistance, $R_0= \rho \frac{L}{A}$, of the microchannel of length $L$, cross-sectional area $A$ and filled with a solution of resistivity $\rho$ can  be estimated  as \cite{weatherall2015applications}
\begin{equation}
R_0 \approx \rho \frac{\pi \left(r_{om}+r_{im}\right)}{8 h (r_{om}-r_{im}) } 
\label{eq_2_} 
\end{equation}
The forces on the walls perturb the channel inner radius to $r_{im}+\frac{2}{3} \Delta r$, which results in both a length change and a cross-sectional area change.  This can be expressed  as 
\begin{equation}
{R_0} + \Delta R \approx \rho \frac{\pi \left(r_{im} +r_{om} + \frac{2}{3} \Delta r \right)}{8 h (r_{om}-r_{im}+\frac{2}{3} \Delta r) }. 
\label{eq_3_} 
\end{equation}
This expression can be approximated as   
\begin{equation}
{R_0} + \Delta R \approx \rho {{\pi \left( {{r_{om}} + {r_{im}} + {2 \over 3}\Delta r} \right)} \over {8h({r_{om}} - {r_{im}})}}{\rm{ }}\left( {1 - {{{2 \over 3}\Delta r} \over {{r_{om}} - {r_{im}}}}} \right) \approx \rho {\pi  \over {8h}}\left[ {{{\left( {{r_{om}} + {r_{im}}} \right)} \over {({r_{om}} - {r_{im}})}} + {{2\Delta r} \over {3({r_{om}} - {r_{im}})}}} \right]\left[ {1 - {{2\Delta r} \over {3({r_{om}} - {r_{im}})}}} \right]
\label{eq_4_} 
\end{equation}
Keeping terms up to order two in $\Delta r$, we find the required expression for the resistance change $\Delta R$ as

\begin{equation}
{{\Delta R} \over {{R_0}}} \approx  - {4 \over 3}\left[ {{{{r_{im}}\Delta r} \over {({r_{om}}^2 - {r_{im}}^2)}} + {{{{\left( {\Delta r} \right)}^2}} \over {3({r_{om}}^2 - {r_{im}}^2)}}} \right]
\label{eq_7_} 
\end{equation}

\begin{figure}[b]
    \centering
    \includegraphics[width=0.9\linewidth]{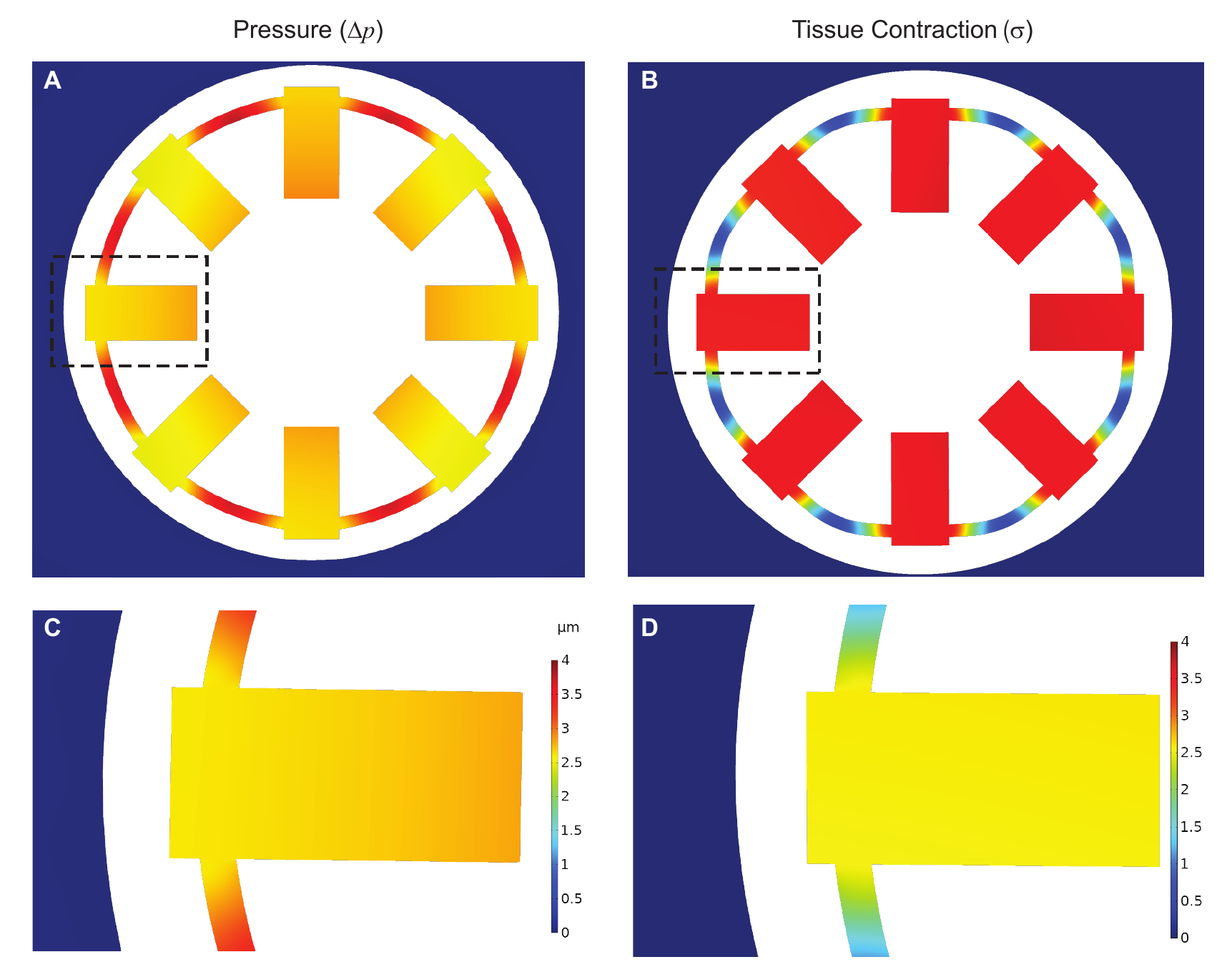}
    \caption{Top view of simulated bending patterns of a 20 $\mu$m-thick cylindrical shell at $z=h/2$ upon applied external pressure (A,C) and tissue contractions (B,D). Top images show overall bending patterns that correspond to same applied force ($\approx 50 ~\rm \mu$N) from outer walls (A) and stubs of the cages (B), where colormap represents normalized amplitude of shell displacement. C and D display deformations experienced at the sensing site, as a response to pressure applied from outer walls ($\Delta p = 2 ~\rm kPa$) and stubs of the cages ($\sigma = 6 ~\rm kPa$) respectively. Even though cage displacement in both cases are approximately the same, $\Delta r(z=h/2) \approx 3 ~\rm \mu m$, $\overline{\Delta r(\theta )_{\Delta p}} > \overline{\Delta r(\theta )_{\sigma}}$.}
    \label{fig:comsol2}
\end{figure}

Eq. \ref{eq_7_} captures the  relationship between  $\Delta R/R_0$ and $\Delta r$. Using the experimental linear calibration relationship between $\Delta r$ and $\Delta p$ in Eq. \ref{eq_7_}, we can estimate, as a check, the  electrical responsivity ${\cal R}_e = \frac {\partial \Delta R}{R_0 \partial \Delta p}$  for the 20-µm-thick and 30-µm-thick devices. Using the linear dimensions of the devices, we found  ${\cal R}_e\approx -1.65 \times 10^{-5} ~\rm Pa^{-1}$ for the 20-µm-thick device and ${\cal R}_e\approx -8 \times 10^{-6} ~\rm Pa^{-1}$ for the 30-µm-thick device. The theoretical ${\cal R}_e$ values found from Eq. \ref{eq_7_} are roughly $ 2.5 \times$ larger than the experimental ${\cal R}_e$ values reported in Table 1 in main text. We suspect that this discrepancy is due to the parasitic contributions to the experimental value of $R_0$. The experimentally measured value of $R_0$ is also approximately a factor of 2.5 larger than the theoretically estimated resistance of  the sensing region based on Eq. \ref{eq_2_}. Residual contact resistances  increase the $R_0$, which lowers the relative resistance change $\Delta R/R_0$ and hence  ${\cal R}_e$. 

Lastly, we took a closer look at  simulations to investigate the effect of different bending patterns resulting from (1) $\Delta p$ applied on the outer wall of the shell and (2)  stress  $\sigma$ exerted on the cage microstructures. Fig. \ref{fig:comsol2} shows the simulated bending patterns of a 20-$\mu$m device from top at $z=h_{m}/2$, when the same force of magnitude $\approx$ 50 $\mu$N is applied as an external pressure (Fig. \ref{fig:comsol2}A) and as tissue contractions from each cage stub (Fig. \ref{fig:comsol2}B). These simulations show that the deformation profile is not uniform across the circumference of the shell at $z=h_{m}/2$.  Since cage displacement $\Delta r$ is the critical parameter for  calibration, we  focus  on the sensing region and cage displacements. In Fig. \ref{fig:comsol2}C and D, we simulated two cases where cage displacements are approximately the same $\Delta r\approx 2.5-3 ~\rm \mu m$ with $\Delta p = 2 ~\rm kPa$ and $\sigma = 6 ~\rm kPa$ respectively. Even though the approximate cage displacements are the same for these two cases, the average displacements across the sensing region are different because of the complex $\theta$ dependence of $\Delta r(\theta )$.  This clearly shows the limitation of the  theoretical expressions derived above using the assumption of $\theta$ independence.  We observed this difference experimentally when analyzing our  electrical signals and when comparing forces exerted simultaneously both from the inside and outside of the seeding well (i.e., Fig. 5 in the main text). In these cases, we corrected the signals empirically, based on optical cage displacements.

 \subsection{{Details of Electrical Resistance Measurement}}
 
\begin{figure}[h!]
    \includegraphics{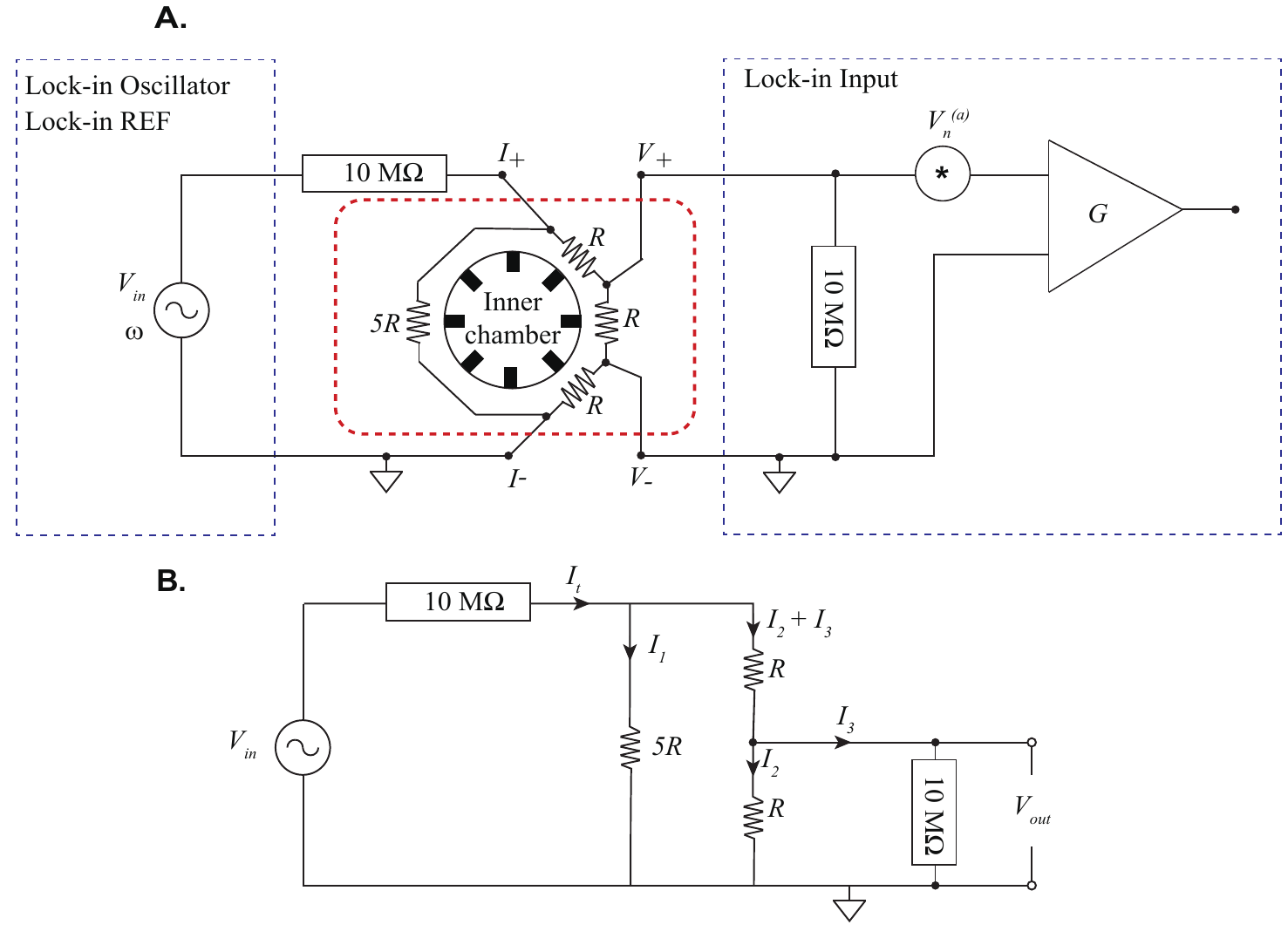}
    \caption{Equivalent electrical circuit for the four wire resistive monitoring scheme. A: $R$ is the electrical resistance of the outer microchannel that is being measured, it is 1/8$^{th}$ of the annular microchannel. The blue dashed boxes represent the lock-in amplifier; $V_{in}= 5 ~\rm V$, $\omega=25 ~\rm Hz$, $V_{n}^{(a)}$ is the input noise voltage and $G$ is the gain of the lock-in amplifier. Red dashed box represents the device. Approximately 500 nA AC current ($I_{RMS}$) goes through the device, and voltage drop on a microchannel that is adjacent to an attachment site is monitored locally for the displacement and force sensing. In measurement of a single device, approximately 40$\%$ of the injected current strays and loops around the annular microchannel. B: A simplified electrical circuit of a single device in order to estimate $R$ from the measured voltage $V_{out}$.}
    \label{fig:circuit_single}
\end{figure}

Electrical resistance  of the microchannel is monitored based on a four-wire measurement scheme using a lock-in amplifier (SR830, Stanford Research Systems). Circuit diagrams are shown  in Figure \ref{fig:circuit_single}.   

Following Fig. \ref{fig:circuit_single}B, we calculated the value of $R$ from the measured voltage $V_{out}$ by using basic circuit analysis:
\begin{equation}
R\approx \frac{{{V}}_{{out}}}{I_2}=\ \frac{{{V}}_{{out}}}{I_t-I_1-I_3}
\label{eq_8_}
\end{equation}
Since $R\ll 10~\rm M\Omega$, the input current $I_t$ can be approximated as $I_t\approx \frac{V_{in}-2V_{out}}{10~\rm M\Omega}$. Likewise, $I_1\approx \frac{2V_{out}}{5 R}$ and $I_3=\frac{V_{out}}{10~\rm M\Omega}$. Combining all,  $R$ is approximately found as
\begin{equation}
R\approx \frac{7V_{out}}{5V_{in}-15V_{out}}\times{10}^7 ~\rm \Omega
\label{eq_11_}
\end{equation}

\begin{figure}[h!]
    \includegraphics{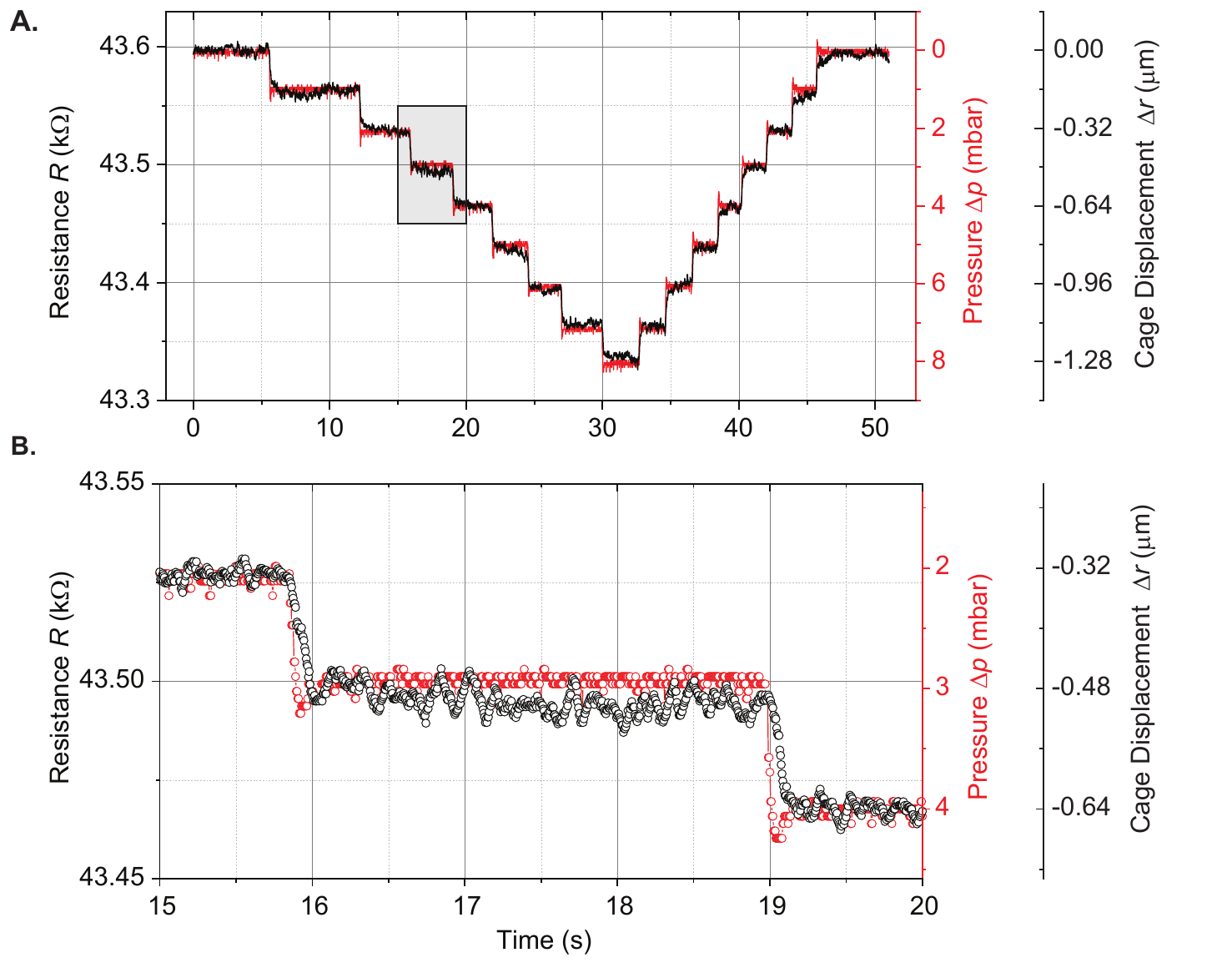}
    \caption{Resolution of pressure regulation and resistance measurement from a device with 20 µm thick cylindrical shell. A: Resistance $R$ (black) is tracked while pressure $\Delta p$ (red) is varied between 0-8 mbar with 1 mbar increments. B: Close-up view to the highlighted area in A, where two 1 mbar steps are shown.}
    \label{fig:resolution}
\end{figure}
 
 \subsection{Estimation of Sensitivity Limits}

Here, we briefly describe how we estimated the various sensitivity limits for the device.  In order to to estimate the  limits of resistance measurement, we determined the root-mean square (rms) resistance fluctuations normalized by the mean resistance $R_0$ value when the system was under equilibrium.  In Fig. \ref{fig:resolution}A, the equilibrium regions are on the plateaus. The rms value  of normalized resistance fluctuations then gave us a minimum detectable resistance shift of   $(\Delta R/R_0)_{min} \approx 6\times {10}^{-5}$ at a signal-to-noise ratio of 1. The equivalent bandwidth here was 15 Hz, which  allows us to estimate a noise floor of $0.67 ~\rm \Omega /Hz^{1/2}$. Note that by measuring longer, the sensitivity may be further improved but the value $0.67 ~\Omega/\rm Hz^{1/2}$ establishes a helpful baseline. We estimate that our sensitivity is limited by the input noise of the lock-in, $V_{n}^{(a)}$, at 20$\times$ gain ($\approx 100 ~\rm nV/Hz^{1/2}$) and the Johnson noise of the resistor ($\approx 30 ~\rm nV/Hz^{1/2}$). Combining these two noise sources with $V_{out}$ yield a theoretical noise floor of $\approx 0.3~\rm \Omega /Hz^{1/2}$, which is not far from the experimentally measured noise floor.

By using the relation of ${\cal R}_e = \frac {\partial \Delta R}{R_0 \partial \Delta p}$, we converted the minimum detectable resistance to a minimum detectable pressure $\Delta p_{min}$. We estimated that $\Delta p_{min}\approx 8.8 ~\rm Pa/Hz^{1/2}$ for the 20-$\mu$m-thick planar device and $\Delta p_{min}\approx 12.8 ~\rm Pa/Hz^{1/2}$ for 30-$\mu$m-thick curved device. Next, we converted  $\Delta p_{min}$ to a minimum detectable cage displacement, $\Delta r_{min}$, by using the linear relation between $\Delta r$ and $\Delta p$ (see Fig. 2B in the main text). Lastly, $\Delta r_{min}$ is converted to a minimum detectable force by $F=k_{eff} \Delta r$.  The noise limits for 20 $\mu$m-thick planar and 30 $\mu$m-thick curved devices are summarized in Table 1 in the main text.   

 \subsection{Available Bandwidth}
 
 The available bandwidth of the device can be inferred from the pulse excitation shown in Fig. 5A in the main text and  Fig. \ref{fig:resolution}B. The measured rise time (or the decay time) of $\tau \sim 0.1~\rm s$ can be converted to a bandwidth as $BW \approx\frac{0.35}{\tau}$. In Fig. \ref{fig:resolution}B, the pressure readout from the sensor embedded in the piezoelectric micropump is tracked simultaneously with the electrical signal (red data trace). The pump outputs a step in a time scale of $\sim 20 ~\rm ms$.  Our analysis below suggests that the overall bandwidth of the system is probably limited by the intrinsic mechanical properties of the PDMS shells.

  \begin{figure}[h!]
     \centering
     \includegraphics{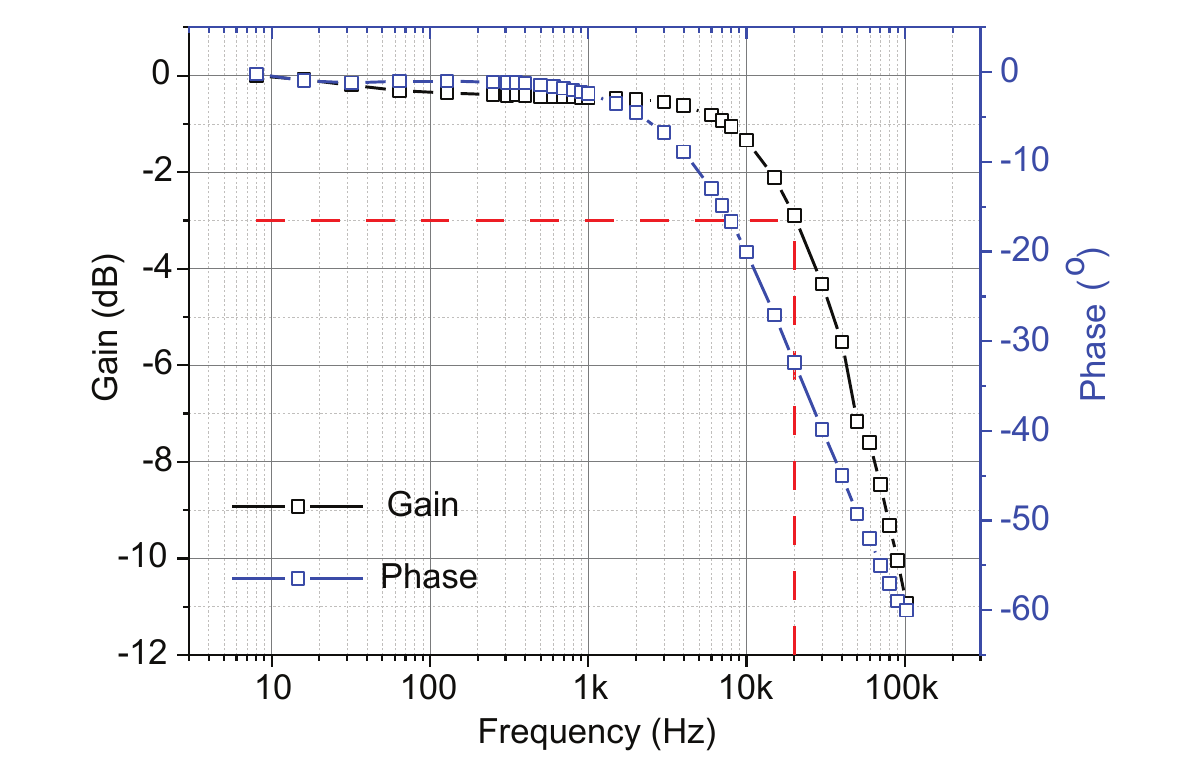}
     \caption{Frequency response of a single device with 0.65 mm diameter Ag/AgCl wire electrodes. Output voltage corresponding to $R$ is measured under same experimental conditions that is used in this study but different driving frequencies. Red dashed lines represent -3dB cutoff point, corresponding to a cutoff frequency of $\sim$ 20 kHz.}
     \label{fig:frequencyresponse}
 \end{figure}
 
 \subsubsection{Electrical Bandwidth}

Fig. \ref{fig:frequencyresponse} shows the frequency response of a the electrical readout circuit.  The electrical circuit  in Fig. \ref{fig:circuit_single} is used for measurement, with the carrier frequency  swept between 8 Hz - 100 kHz. Fig. \ref{fig:frequencyresponse} displays the magnitude  (normalized to its low-frequency value in units of dB)  and  the phase shift of the output as a function of  frequency. We observe that the cutoff frequency is $f_c \approx 20 ~\rm kHz$, which corresponds to a time constant of $\tau \approx 8 ~\rm \mu s$. This is the maximum available electrical bandwidth for the device and circuit in this study. Note that we took advantage of this  electrical bandwidth when picking four different carrier frequencies $\gtrsim 200 ~\rm Hz$ during the parallel sensing endeavour.  It is also worth emphasizing that we typically did not use the full bandwidth, instead we optimized the noise performance by using a lock-in time constant of $ \sim 3-10 ~\rm ms$.  This bandwidth is still significantly larger than the observed bandwidth, indicating that the system is not limited by the response time of the electrical circuit. 
 
  \subsubsection{Mechanical Bandwidth}
  
Rise time of the pressure pulse applied by the pump appears to be $\sim$ 10-20 ms and  is considerably faster than the observed mechanical response (Fig \ref{fig:resolution}B, red curve vs. black curve). Assuming that there is negligible fluid flow in the system during the actuation and detection,  pressure waves should propagate at the speed of sound. This should not cause a delay between the applied pressure and the observed mechanical response. Thus, it seems probable that the stress relaxation time of the PDMS limits the response time of the system \cite{dogru2018poisson}. Indeed, the observed mechanical bandwidth of the system  in Fig \ref{fig:resolution} is on the same order with the reported relaxation times of other PDMS membrane based  pneumatic actuators \cite{flexcell}.
 
 \subsection{High-throughput Contractility Measurement and Electrical Cross-talk}
 
\begin{figure}[h!]
     \centering
     \includegraphics{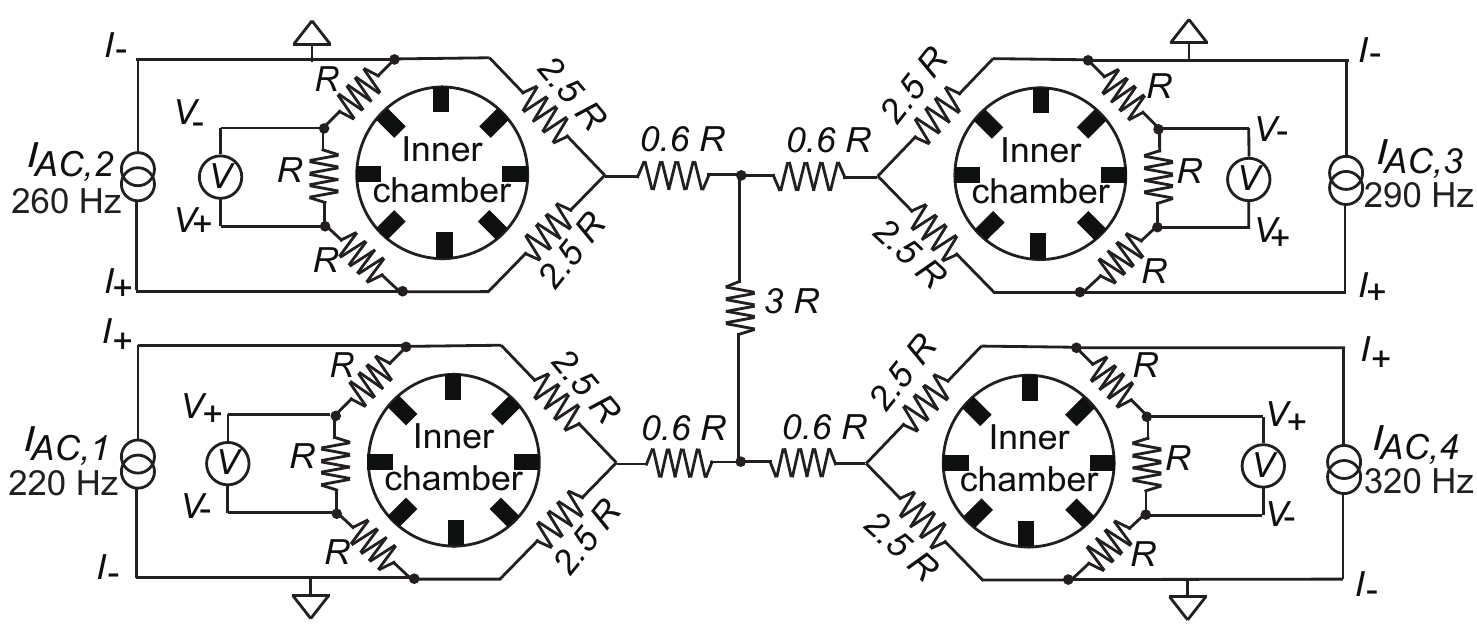}
     \caption{Electrical circuit diagram of the overall platform and the experimental setup to measure contractile forces from all four devices simultaneously.}
     \label{fig:crosstalk}
 \end{figure}

Fig. \ref{fig:crosstalk} shows the electrical circuit diagram of the entire platform. Here, we estimate that $\sim 40\%$ of the injected current from each current source  couples to  other devices. Regardless,  it is possible to avoid cross-talk and measure multiple devices in parallel if current is injected at different carrier frequencies and phase sensitive narrowband detection  is employed. By using four custom-built portable lock-in amplifiers and taking advantage of the available electrical bandwidth (Fig. \ref{fig:frequencyresponse}), we performed sensing at frequencies of 220 Hz, 260 Hz, 290 Hz and 320 Hz. We thus measured the active contractions from all four devices inside the platform. The equivalent  bandwidth for these lock-in amplifiers were $\approx 10 ~\rm Hz$, thus the frequency band allocated to each sensor did not overlap with  others.

 \section{Source of Nonlinearities}
 
  \subsection{Mechanical Nonlinearity}
 
Based on the experimental results, we  made a linear approximation between the applied pressure $\Delta p$ and the  displacement $\Delta r$ (see Fig. 2B in main text) within the pressure range  used in this study. Upon closer inspection,  we  observe that the 30-µm-curved shells are slightly more nonlinear and  hysteretic compared to the  20-µm-planar shells. It appears that the curved geometry is slightly more responsive to the negative pressure than the positive pressure. The mechanical responsivity difference between $\Delta p<0$ and $\Delta p>0$ regimes is also observed in finite element simulations. It is possible to take into account this difference by using a linear fit for each regime at the expense of complexity. At extreme pressure differentials, however, it is clear from Movie S2 ($\Delta p \pm 400 ~\rm mbar$) that stretching occurs as well as bending of the cylindrical shells. Hence, the combined effect of bending and stretching, as well as hyperelastic properties of PDMS, makes the relation between $\Delta p$ and $\Delta r$ nonlinear at the extreme limits.    
 
 \subsection{Electrical Nonlinearity}
 
After plugging in the dimensions for $r_{om}$ and $r_{im}$ and solving  Eq. \ref{eq_7_} with the nonlinear term $(\Delta R/R_0)_{NL}$,  and without the nonlinear term, $(\Delta R/R_0)_{L}$, we observe that the relation between $\Delta R/R_0$ and $\Delta r$ remains linear in the displacement range  $- 10 ~{\rm \mu m} < \Delta r < 10 ~{\rm \mu m} $: $\left \lvert \frac{(\Delta R/R_0)_{NL}-(\Delta R/R_0)_{L}}{(\Delta R/R_0)_{L}}  < 0.01 \right \lvert$. In our calibration and experiments, $\Delta r$ remains in the range $- 8 ~{\rm \mu m} < \Delta r < 8 ~{\rm \mu m} $. The tissue generated displacements were even smaller, $<4 ~{\rm \mu m}$, making it acceptable to  use a linear relation between $\Delta R/R_0$ and $\Delta r$. However, the resistance change also becomes nonlinear in extreme cases (such as that shown in Movie S2, where $\Delta r \approx \pm 60 ~\rm \mu m$).

 \section{Supplementary Movies}

 \subsection{Movie S1}

Response of a cage to an externally applied triangular pressure waveform  with a period of 5 s  and approximate amplitude of $ \pm 50~\rm mbar$ . 

 \subsection{Movie S2}

Maximum strains achievable with the platform. Actuation of the device with a 30-µm-curved shell with a triangular pressure waveform with a period of 10 s and amplitude of $\pm 400$ mbar. The shell is still durable and isolates the inner chamber from outer the chamber. 

 \subsection{Movie S3}

Attachment sites provide physical cues to define the geometry of the engineered cardiac microtissues. Spontaneous contractions from devices with attachment sites that are arranged in octagonal (left), pentagonal (middle) and rectangular (right) configurations.

  \subsection{Movie S4}

Spontaneous contractions of cardiac microtissues in a 20-µm-thick planar (left) and a 30-µm-thick curved (right) devices, within the same platform.

  \subsection{Movie S5}

Mechanical pacing of a microtissue in a device with a 20-µm-thick planar shell. Approximately 1.1\% tensile strain is applied with the pump ($\Delta p-20$ mbar, 250 ms long square pulses with 0.5 Hz frequency), while the engineered cardiac microtissue  spontaneously beats at a frequency around 0.25 Hz, resulting in an approximately  0.75\% compressive strain.
 
  \subsection{Movie S6}

 Spontaneous contractions of cardiac microtissues in all four devices  on the same platform (left: 30-µm-thick curved devices, right: 20-µm-thick planar devices).

\clearpage

\bibliography{supplementary.bbl}